 \documentclass[11pt,preprint]{aastex}

\usepackage{graphicx,amssymb,natbib}

\newcommand{\Teff}{$T_{\rm eff}$}
\newcommand{\logg}{$\log{g}$}
\newcommand{\vmicro}{$v_{\rm micro}$}
\newcommand{\kms}{km\,s$^{-1}$}
\newcommand{\degrees}{$^{\circ}$}
\newcommand{\Msun}{$M_{\odot}$}
\newcommand{\Me}{$M_{\oplus}$}
\defcitealias{Gust:75}{GBEN75}

\slugcomment{AJ, accepted 11 July 2003}

\shorttitle{Abundance Analysis of Planetary Host Stars}

\begin{document}

\title{Abundance Analysis of Planetary Host Stars\\
       I. Differential Iron Abundances}

\author{U. Heiter and R. E. Luck}
\affil{Department of Astronomy, Case Western Reserve University, Cleveland, OH 44106}
\email{ulrike@fafnir.astr.cwru.edu, luck@fafnir.astr.cwru.edu}

\begin{abstract}
We present atmospheric parameters and iron abundances derived from high-resolution spectra for three samples of dwarf stars: stars which are known to host close-in giant planets (CGP), stars for which radial velocity data exclude the presence of a close-in giant planetary companion (no-CGP), as well as a random sample of dwarfs with a spectral type and magnitude distribution similar to that of the planetary host stars (control). All stars have been observed with the same instrument and have been analyzed using the same model atmospheres, atomic data and equivalent width modeling program. Abundances have been derived differentially to the Sun, using a solar spectrum obtained with Callisto as the reflector with the same instrumentation.

We find that
the iron abundances of CGP dwarfs are on average by 0.22~dex greater than that of no-CGP dwarfs.
The iron abundance distributions of both the CGP and no-CGP dwarfs are different than that of the control dwarfs, while the combined iron abundances have a distribution which is very similar to that of the control dwarfs.
All four samples (CGP, no-CGP, combined, control) have different effective temperature distributions.
We show that metal enrichment occurs only for CGP dwarfs with temperatures just below solar and $\approx$ 300~K higher than solar, whereas the abundance difference is insignificant at \Teff\ around 6000~K.
\end{abstract}


\section{Introduction}
\label{introduction}

The metallicity of planetary host stars has been the subject of several studies carried out in recent years, including abundance analyses based on high-resolution spectra \citep[e.g.][]{Sada:02,Sant:03,Laws:03}. Iron abundance distributions for planet hosts have been compared to that of field star samples, for which the abundances had been derived by different methods, or a sample of stars ``without detected giant planets'' \citep[ abundances derived with same method as used for planet hosts]{Sant:01a}. A common result of these studies is that the metallicity distributions of planet host and comparison samples are not the same. The main difference that has been found is that the average iron abundances of the planet hosts are about 0.2~dex higher than that of the comparison samples.

The work presented in the current paper was conducted as part of a high resolution spectroscopic survey of local neighborhood stars.  This program seeks information on the population of stars in the local neighborhood, i.e., those stars within several hundred parsecs of the Sun. 
The parent sample for this program consisted of those stars in the Hipparcos Parallax Catalogue \citep{Perr:97} which are within 100~pc of the Sun. The following subsets of stars were selected for observation:
\begin{itemize}
\item The stars with distances less than 15~pc with absolute magnitudes brighter than 10.0.
\item Samples of stars with characteristics that make them of interest in regard to planets regardless of distance (see Section~\ref{selection}).
\item A sample of stars within a radius of 100~pc drawn from the Hipparcos Parallax Catalogue. The volume has been subdivided into 25~pc cubes centered at the solar position. Approximately 300 of the sub elements have Hipparcos stars within them. For each sub element we have selected a giant and a dwarf for observation, preferring spectral types G5 to K0 for the giants and G0 to G5 for the dwarfs.
\end{itemize}
In order to obtain a highly consistent dataset, we decided to use a single telescope/instrument combination with the same settings for every star. As the observations are carried out at McDonald observatory, this means that all stars are north of $-$30\degrees\ declination. The first and last set of stars listed above therefore contain 156 and 598 stars, respectively. The second set is described in detail in Section~\ref{selection}.
In this paper we focus on comparing truly complementary samples of stars with regard to their (either known or undetectable) planetary systems. We used a consistent set of observations (see Section~\ref{observations}) and analysis methods (Section~\ref{analysis}) for all stars. Analyzing the stellar data strictly differentially with respect to the Sun enables us to make use of a maximum fraction of the data that have been obtained. In Section~\ref{discussion} we describe the physical properties of the stars in our datasets, inferred from our abundance analysis results, as well as the possible implications for giant planet formation.

\section{Sample Selection}
\label{selection}

Since the announcement of the first candidate planet around a solar-type star \citep{Mayo:95}, the number of publications on the discovery of extrasolar planetary systems has continuously grown. A convenient tool is provided on the web in the form of the ``Extrasolar Planets Encyclopaedia''\footnote{\url{http://www.obspm.fr/encycl/encycl.html}}, which lists on one page all planetary systems announced so far, includes links to all relevant information and references and is updated frequently.
Understandably, the situation is less favorable if one is looking for stars that do not host any planets. A correct assessment of the current situation would be that we do not know any such stars because of the limitations of radial velocity searches in terms of detectability of planets with certain orbital elements. These limitations are: the best achievable radial velocity accuracy \citep[on the order of 1~ms$^{-1}$, see, e.g.,][]{Endl:03}, the time resolution employed for each star, and the maximum time span since the start of the planet search programs. These imply that for every observed star, the existence of planets below a certain minimum mass, depending on the semi-major axis $a$ of the system and the stellar mass, cannot be excluded. Addressing and quantifying these limits is time-consuming and the only results published up to now are \citet{Walk:95,Cumm:99,Endl:02}, based on CFHT, Lick and ESO planet searches, respectively. 

In order to assess the influence of metallicity on the properties of planetary systems, one can basically take one of two different paths: either comparing the stars with planets to a large sample of stars for which the planet status is in general not known, but which reflects the estimated frequency of planetary systems, or comparing the element abundances of stellar groups which are as complementary as possible in regard to the properties of their planets while being also of approximately the same size. We chose the latter approach to select stars for a detailed abundance analysis and consequently excluded a number of stars both from the planet-host list and from the lists of stars with published upper mass limits. However, in Section~\ref{stat}, we also discuss the results of taking the first approach using photometric calibrations.

The selection is based on the linear relation between the logarithmic minimum planet mass and the logarithmic orbital radius for a circular orbit with planet mass $M_{\rm p}$ much less than stellar mass $M_{\ast}$, with slope 1/2. We compiled the following samples of stars using the given selection criteria ($M_{\rm p} \sin i$ in Jupiter masses, $a$ in AU, $i$ is the orbital inclination): 

\begin{enumerate}
\item ``CGP'' dwarfs -- close-in giant planet hosts listed in the Extrasolar Planets Encyclopaedia. This excludes systems with $M_{\rm p} \sin i$ greater than an adopted brown dwarf mass limit of 13~$M_{\rm J}$. From this list we selected stars with $\log (M_{\rm p} \sin i) > 0.5 \log a + 0.1$. In addition, stars hosting planets with eccentricity $e > 0.5$ were excluded, because tests with non-circular orbits suggest that planet detectability decreases with increasing $e$ \citep{Endl:02}. 
\item ``no-CGP'' dwarfs  -- stars included in planet searches for which upper mass limits for detectable planets have been published. Only stars with $\log (M_{\rm p} \sin i)_{\rm max} \le 0.5 \log a + 0.1$ were selected.
\item ``control'' dwarfs  -- stars randomly selected from the Bright Star Catalogue of similar spectral type to those of 1 and 2. Ideally, the properties of this sample should be identical to that obtained by combining samples 1 and 2. 
\end{enumerate}

We further excluded two stars from the CGP dwarf sample which were searched for planets based on their high metallicities (see Appendix).
With the restriction to declinations north of $-$30\degrees, sample 1 contains 34 stars, sample 2 35 stars, and sample 3 32 stars.

\begin{figure}
\resizebox{\hsize}{!}{ \includegraphics{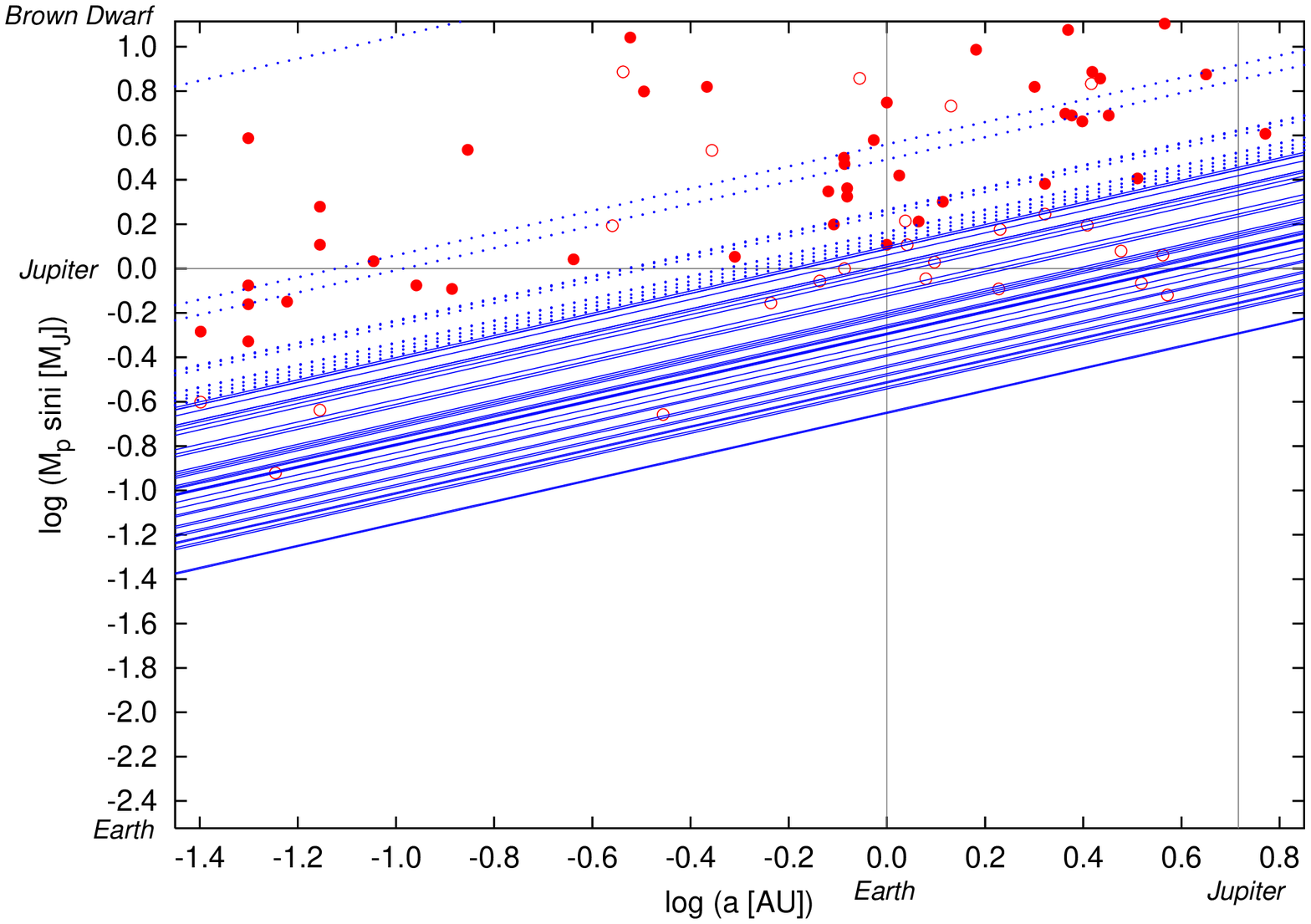} }
\caption{Red dots: minimum masses and semi-major axes for CGP dwarfs from the Extrasolar Planet Encyclopaedia. Blue lines: upper limits for minimum planet masses as a function of orbital radius ($e$=0), derived from mean upper limits on radial velocity amplitudes from \citet{Walk:95,Cumm:99,Endl:02}, defining the no-CGP dwarf sample. Open circles and dotted lines correspond to planet properties and upper limits for excluded stars.}
\label{upper_limits}
\end{figure}

The orbital properties of the planetary systems that make up sample 1 and can (not) exist in sample 2 are illustrated in Figure~\ref{upper_limits}. The figure shows the upper limits for minimum planet masses as a function of orbital radius, derived from published mean upper limits on radial velocity amplitudes (blue lines), and minimum masses and semi-major axes for known extrasolar planets (red dots). By applying the selection criteria we excluded all stars hosting planets with properties shown by the open circles and (no-planet) stars with detection limits corresponding to the dotted lines.

Even though we decided to limit the discussion of our results to the samples selected as described above, we observed and analyzed almost all planet hosts known up to now with declinations $>-30$\degrees. The results are given in the appendix.

\subsection{Very Strong Lined Dwarfs}

A fourth group of stars has been observed, which we call ``Very Strong Lined'' or VSL dwarfs.
These are the 23 (northern) stars listed in Table 5 of \citet{Egge:78}, as well as HD\,145675 \citep{Tayl:70}, a well-known strong lined star. \citet{Egge:78} selected these stars based on enhanced metallicities ([Fe/H] $\ge$ +0.1) derived from a calibration of photometric indices in a modified Str\"omgren system and spectroscopic abundances determined for a few stars in several moving groups by various authors in the 1960s and '70s. Stars similar to these had been studied previously using a photoelectric spectrum scanner technique for determining abundances and had been designated "super-metal-rich" (SMR) by \citet[ giants]{Spin:69} and \citet[ dwarfs]{Tayl:70}.
Recently, \citet{Luck:95} conducted an abundance study of 33 VSL field giants and compared them to normal giant stars. \citet{Felt:98} and \citet{Felt:01} analyzed metal rich dwarf stars.
We obtained high resolution spectra in order to examine if we obtain overabundances for the dwarfs ``classified'' as VSL. If this is the case, it would be interesting to know if these stars do or do not host planets.
It turned out that, from our list of VSL dwarfs, four are already included in the CGP dwarf list, four are no-CGP dwarfs and for three further stars (HR\,166, HR\,508 and HR\,3626) our spectra indicate that they might be double-lined spectroscopic binaries. This leaves 13 additional VSL dwarfs that have been analyzed in the present work.

\section{Observations}
\label{observations}

High signal-to-noise spectra were obtained during several observing runs between 1997 and 2002 (see Tables~\ref{obs_log_dwp} to \ref{obs_log_vsl}).
For all observations we used the Sandiford Cassegrain Echelle Spectrograph \citep{McCa:93} attached to the 2.1m telescope at McDonald Observatory.
The spectra continuously cover a wavelength range from about 4840 to 7000\,\AA\ with a resolving power of about 60000. 
Each night we observed also a broad lined B star with S/N several times exceeding that of the program stars, to enable cancellation of telluric lines where necessary.

\begin{figure}
\resizebox{\hsize}{!}{ \includegraphics{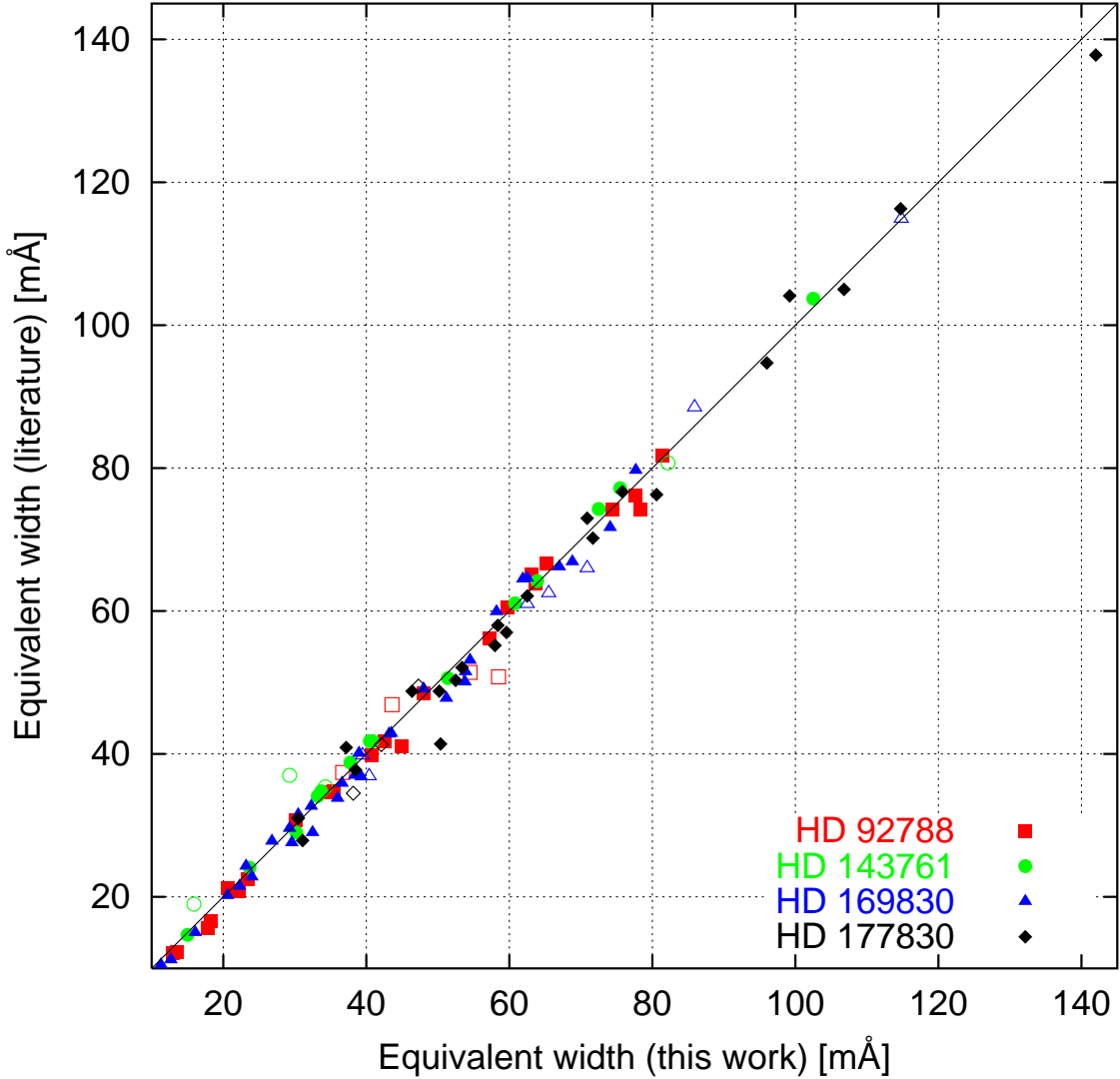} }
\caption{Comparison between equivalent widths determined in this work and by the following authors: HD\,92788: \citet{Gonz:01} (average difference for 27 lines: +2.9$\pm$5.5\%), HD\,143761 ($\rho$ CrB): \citet{Gonz:98a} (18 lines: $-$2.9$\pm$6.1\%), HD\,169830: \citet{Sant:00} (39 lines: +2.5$\pm$4.7\%), HD\,177830: \citet{Gonz:01} (25 lines: +2.1$\pm$6.2\%). Full and open symbols correspond to Fe~I and Fe~II lines, respectively.}
\label{ew_dwp}
\end{figure}

We used IRAF\footnote{IRAF is distributed by the National Optical Astronomy Observatories, which are operated by the Association of Universities for Research in Astronomy, Inc., under cooperative agreement with the National Science Foundation.} to perform CCD processing, scattered light subtraction, and echelle order extraction.
For all further reductions a Windows based graphical package (ASP) developed by REL was used \citep[see, e.g.,][]{Luck:94}. This included Beer's law removal of telluric lines, smoothing with a fast Fourier transform procedure, continuum normalization, and wavelength calibration using template spectra. Finally, equivalent widths ($W$) were determined using the Gaussian approximation to the line profile. For lines with multiple measurements average fractional differences in $W$ are in general lower than 15\% for 10~m\AA~$< W <$ 20~m\AA, lower than 10\% for 20~m\AA~$< W <$ 30~m\AA, and lower than 5\% for $W >$ 30~m\AA. A comparison with published values indicates agreement at the 3\% level for most stars (see Fig.~\ref{ew_dwp}).
Only lines with equivalent widths between 10 and 200~m\AA\ were used for the analysis.

Listings of the observations including central wavelength and S/N for each echelle order can be obtained from the department's research webpage\footnote{\url{http://burro.astr.cwru.edu/dept/research.html\#resources}}. This site also provides access to the data themselves, ie. the extracted spectra in raw, edited or normalized form. Data from the other two sets of stars mentioned in Section~\ref{introduction} will also be made available.

\begin{figure}
\resizebox{\hsize}{!}{ \includegraphics{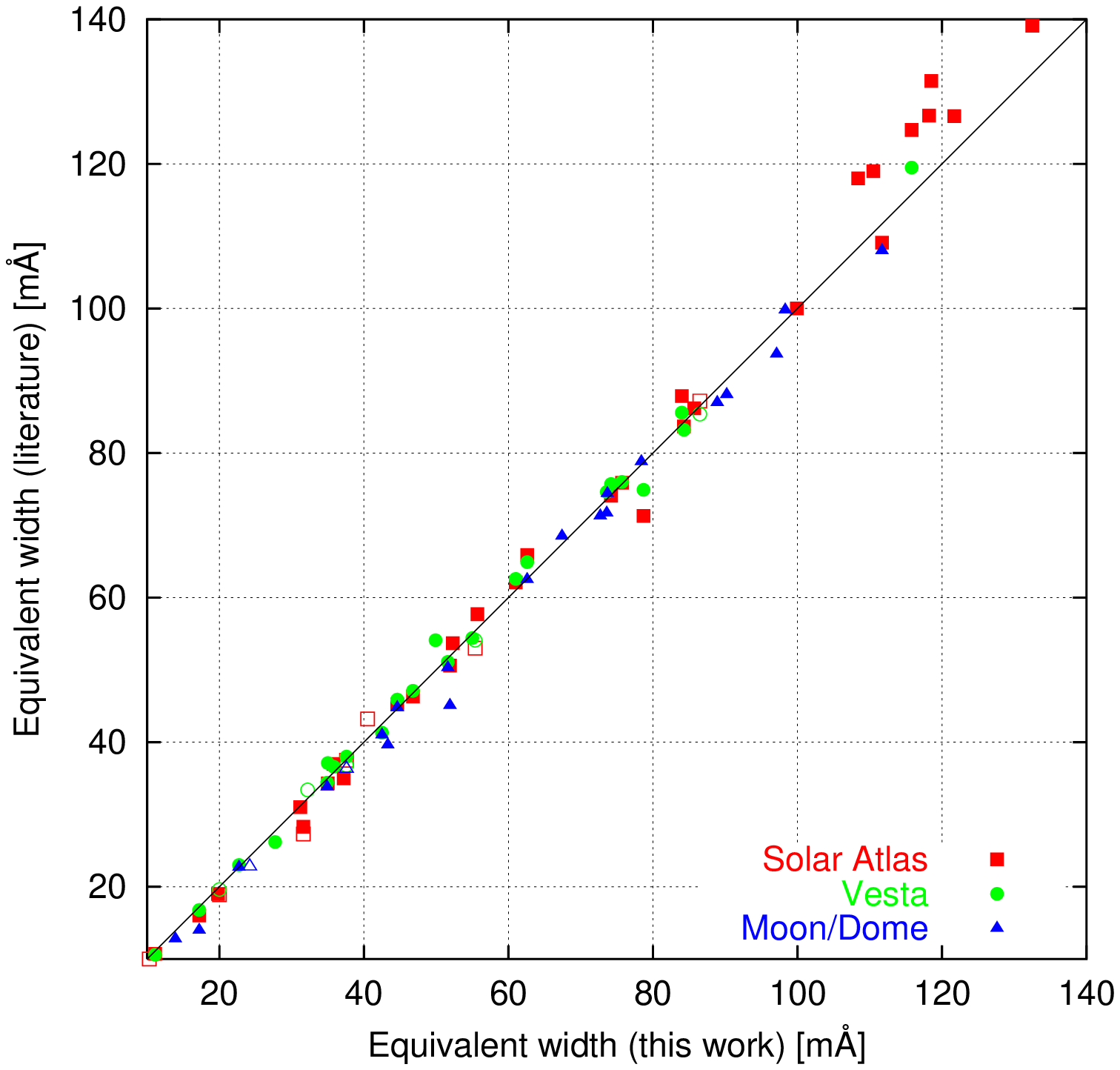} }
\caption{Comparison between solar equivalent widths determined in this work and by the following authors: ``Solar Atlas'': \citet{Gonz:96} and \citet{Gonz:98a} measured equivalent widths on the Solar Flux Atlas \citep{Kuru:84} (average difference for 37 lines: +0.1$\pm$5.7\%), ``Vesta'': \citet{Gonz:97} and \citet{Gonz:98a} used a spectrum of the Sun, reflected off the asteroid Vesta (28 lines: $-$0.2$\pm$3.2\%), ``Moon/Dome'': \citet{Edva:93b} used observations of the Moon or the sun-lit white dome of the ESO 3.6m telescope (24 lines: +4.8$\pm$7.3\%). Full and open symbols correspond to Fe~I and Fe~II lines, respectively.}
\label{ew_sun}
\end{figure}

\begin{figure}
\resizebox{\hsize}{!}{ \includegraphics{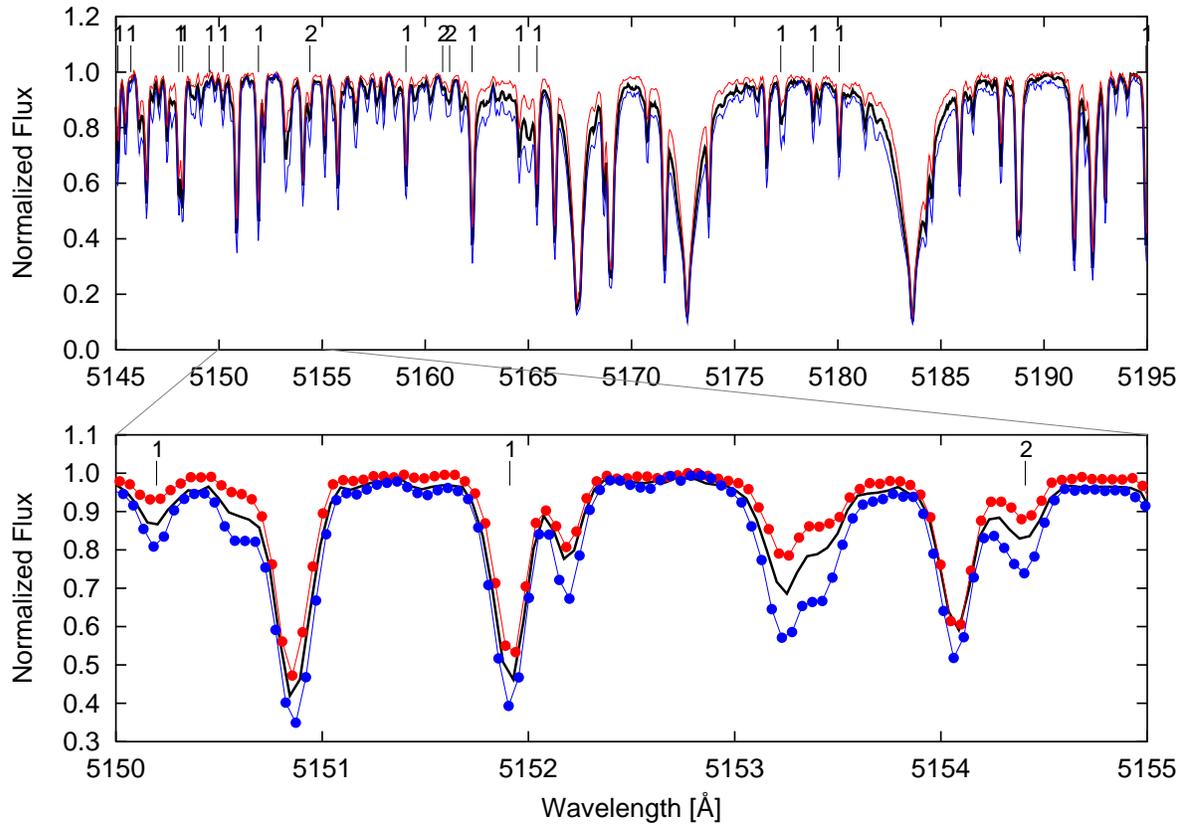} }
\caption{Sample spectra: Sun (Callisto, black thick line) and two CGP dwarfs: HD\,92788 ([Fe/H]=+0.3, lower blue thin line), HD\,143761 ([Fe/H]=$-$0.25, upper red thin line). The upper panel spans one echelle order, the lower panel shows an enlargement. Measured Fe~I and Fe~II lines are marked with ``1'' and ``2'', respectively.}
\label{observation}
\end{figure}

To enable a differential analysis, we obtained a solar flux spectrum using Callisto as the reflector. We used the same spectrograph and reduction procedure as for our program stars. The measured equivalent widths are in reasonable agreement with that determined by other authors from different sources, as shown in Fig.~\ref{ew_sun}. Figure~\ref{observation} shows a portion of the observed spectra of the Sun and three CGP dwarfs.

\section{Abundance Analysis}
\label{analysis}

After examining published abundance analyses of dwarfs with planets (and related dwarfs) we decided that a mere repetition of the abundance analyses using equivalent data and similar methods would not be profitable. If we use the line selection and atomic data from e.g. \citet{Sant:00} or \citet{Gonz:01} (typically about 30-40 Fe lines) we reproduce their abundance values with high accuracy. If we use a larger selection of iron lines ($\approx$600 lines based on $gf$ values from the NIST Atomic Spectroscopic Database\footnote{\url{http://physics.nist.gov/cgi-bin/AtData/main\_asd}}, supplemented by data from \citet{Thev:89,Thev:90} and \citet{Bard:94} for Fe~I, and on $gf$ values from \citet{Giri:95} for Fe~II) we find a significant scatter in the per line abundances. This means that depending on how the line selection is done one can obtain an [Fe/H] ratio which spans a significant range of values: typical allowed spans are $\pm$0.2 dex. After examining our data we have decided to do a differential abundance analysis relative to the Sun using the larger sample of lines (see Sections~\ref{observations} and \ref{lineabun}). In this way, we avoid having to discard the major part of the observations.

\subsection{Initial Parameters}
\label{init}

The atmospheric parameters \Teff, \logg, \vmicro\ and [Fe/H] were determined from our observations by standard fine analysis techniques. 
We started by estimating initial values for these parameters by one of three methods in the following order of preference: 
\begin{enumerate}
\item Literature data (abundance analyses): \citet{Edva:93a,Fuhr:98}, Gonzalez et al. (1997-2001), \citet{Sant:00,Sant:01b,Sant:01a}
\nocite{Gonz:97,Gonz:98a,Gonz:98b,Gonz:99,Gonz:00,Gonz:01}
\item Calibration for Geneva photometry \citep{Kuen:97} using data from the General Catalogue of Photometric Data \citep{Merm:97}\footnote{\url{http://obswww.unige.ch/gcpd/gcpd.html}}
\item A new calibration for \Teff\ and line-depth ratios for our sample of dwarf stars.
\end{enumerate}

\begin{figure}
\resizebox{\hsize}{!}{ \includegraphics{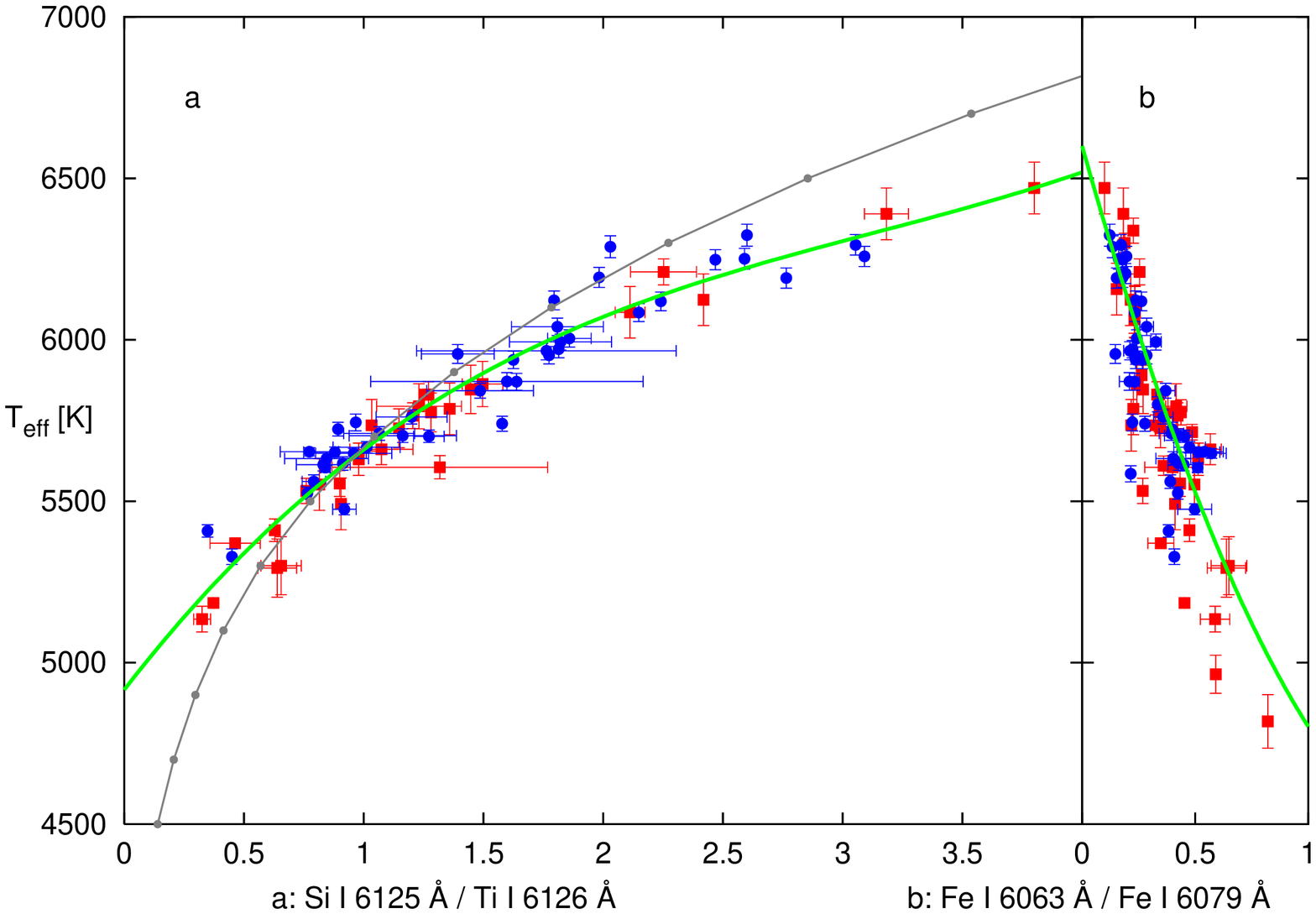} }
\caption{Dependence of line-depth ratios on effective temperature: Squares show spectroscopic temperatures from \citet{Edva:93a,Fuhr:98}, Gonzalez et al. (1997-2001), \citet{Sant:00,Sant:01b,Sant:01a}, circles show temperatures from Geneva photometry \citep{Kuen:97,Merm:97}, the solid line shows a polynomial fit to the data, and the line with points shows ratios calculated from synthetic spectra. A relation for a line-pair useful for temperature calibration is shown in (a), whereas the relation shown in (b) has not been used for the calibration.}
\label{ratios}
\end{figure}

For method 3 we measured the central depths of 45 spectral lines forming the 32 line-pairs which \citet{Kovt:00} had used for a temperature calibration for supergiants. Line-depth ratios for each pair were then plotted against \Teff\ values from methods 1 and 2. We found 20 of these line pairs to be useful for estimating effective temperatures of dwarf stars. As an example, Fig.~\ref{ratios} (a) shows the ratio of the depths of the Si~I 6125\AA\ and Ti~I 6126\AA\ lines as a function of \Teff\ which has been used for the calibration. Fig.~\ref{ratios} (b) shows \Teff\ vs. the depth-ratio of Fe~I 6063\AA\ and Fe~I 6079\AA\ which has not been used for the calibration. The final calibration consists of polynomial fits to the data, or in a few cases, where only few datapoints were available, the relation calculated from synthetic spectra. An updated calibration using the results of the present work will be published in a separate paper.

When \logg\ could not be estimated by methods 1 or 2, it was initially set to a value appropriate for the spectral type of the star.

For each program star we calculated a grid of $5 \times 5$ model atmospheres \citep[ hereafter GBEN75, old version of MARCS code]{Gust:75} centered on the initial parameters, in steps of 100~K and 0.1 for \Teff\ and \logg, respectively. We used opacity distribution functions (ODFs) for scaled solar abundances of $-$1.0, $-$0.5, 0.0 or +0.5~dex, whichever was nearest to the stellar iron abundance.

\subsection{Line Abundances}
\label{lineabun}

Synthetic equivalent widths were calculated for all Fe lines observed for each star (up to about 600 Fe~I and 40 Fe~II lines) for each model atmosphere for 11 \vmicro\ values (steps of 0.1~\kms\ centered on 1.0~\kms) and fit to the observed ones by variation of abundance for each line. We used a modified version of the program LINES \citep[originally][]{Sned:73} and oscillator strength data compiled by REL. We used van der Waals damping constants from \citet{Bark:00}, if available, and the Uns\"old approximation \citep{Unso:38} otherwise. The resulting absolute line abundances were converted to relative line abundances by subtracting the corresponding solar line abundances. These were determined from the Callisto spectrum using a model atmosphere \citepalias{Gust:75} with (\Teff, \logg) = (5777~K, 4.44~[cgs]) and setting \vmicro\ to 0.8~\kms\ \citep{Grev:99}.
When the solar abundance data are used to zero point the program star abundances we find that the standard deviations of the abundances decrease dramatically -- typically by at least a factor of two. The {\em mean} abundances, however, are similar to that obtained when calculating absolute abundances for the stellar and the solar spectrum separately and taking the difference between the two mean abundances. This indicates that there are still major problems in the bulk Fe~I/Fe~II $gf$ value data. This method is analogous to using solar oscillator strengths, but, if desired, we can recover an absolute abundance (with respect to laboratory $gf$ values) with minimum effort.
The differential abundances were used throughout the remaining steps of the analysis.

The numbers of Fe~I lines per 100~\AA\ with relative abundances are on the order of 30 in the region from 4900 to 5900~\AA\ and approximately 15 at longer wavelengths.
For the determination of the best-fit parameters, we used only lines for which the relative abundance deviated from the mean abundances by less than two standard deviations. In this way we largely exclude possible erroneous measurements because of unidentified blends or problems with the continuum setting. This removes about 5\% of the lines from all wavelengths across the whole spectrum and leaves us with about 450 Fe~I and 25 Fe~II lines for each star.
However, for the three coolest stars (HD\,32147, HD\,177830 and HD\,201091, \Teff $\le$ 5000~K) we discarded all lines with wavelengths lower than 5300~\AA\ and selected useful Fe~II lines manually by comparison with synthetic spectra.
For the calculation of the final abundances from the best-fit models we included only lines with abundance deviations less than one standard deviation (about 20\% of the lines with relative abundances removed from all wavelengths), resulting in about 400 Fe~I and 20 Fe~II lines for each star (see Tables~\ref{param_tab_1} to \ref{param_tab_4}). Line data tables are available on request from the authors.

\subsection{Final Parameters and Iron Abundances}
\label{parameters}

The final, best-fit atmospheric parameters were determined by driving the slopes of the relations between equivalent width and differential Fe~I line abundance and between excitation energy and line abundance to zero and demanding ionization equilibrium for the Fe line abundances. The following steps were used to achieve the fulfillment of these criteria
and carried out by a Perl script:

\begin{enumerate}
\item[a] Find the values of \vmicro\ which result in zero slope in the correlations between equivalent width and abundance and between excitation energy and abundance for each of the 25 (\Teff, \logg) models by interpolating in the 11 \vmicro\ values. For most of the (\Teff, \logg) pairs, the two relations give different ``\vmicro\ zero points''.
\item[b] Calculate the set of (\Teff, \logg) values which result in the {\em same} ``\vmicro\ zero point'' value for the equivalent width and excitation energy relations, by linear interpolation (in \logg).
\item[c] Independently from {\em a} and {\em b}, find the set of (\Teff, \logg) models which give the same abundance for Fe~I and Fe~II lines, for the initial (central) \vmicro\ value, again by interpolating (in \logg). The variation of the difference between mean Fe~I and Fe~II abundances with \vmicro\ was negligible for our program stars.
\item[d] Find the model satisfying {\em b and c} by calculating the intersection of the lines connecting the points derived in {\em b} and {\em c}; determine the corresponding \vmicro\ value and iron abundance; round the results to 50~K, 0.05~[cgs] and 0.05~\kms\ for \Teff, \logg\ and \vmicro, respectively; confirm that the derived parameters give both zero slopes and ionization equilibrium.
\item[e] If the derived abundance differed from the ODF abundance by more than 0.25~dex, the procedure was repeated with a grid calculated with the nearest ODF abundance. Furthermore, in some cases the solution was located outside of the initial grid, in which case the grid was extended accordingly.
\end{enumerate}

\begin{figure}
\resizebox{\hsize}{!}{ \includegraphics{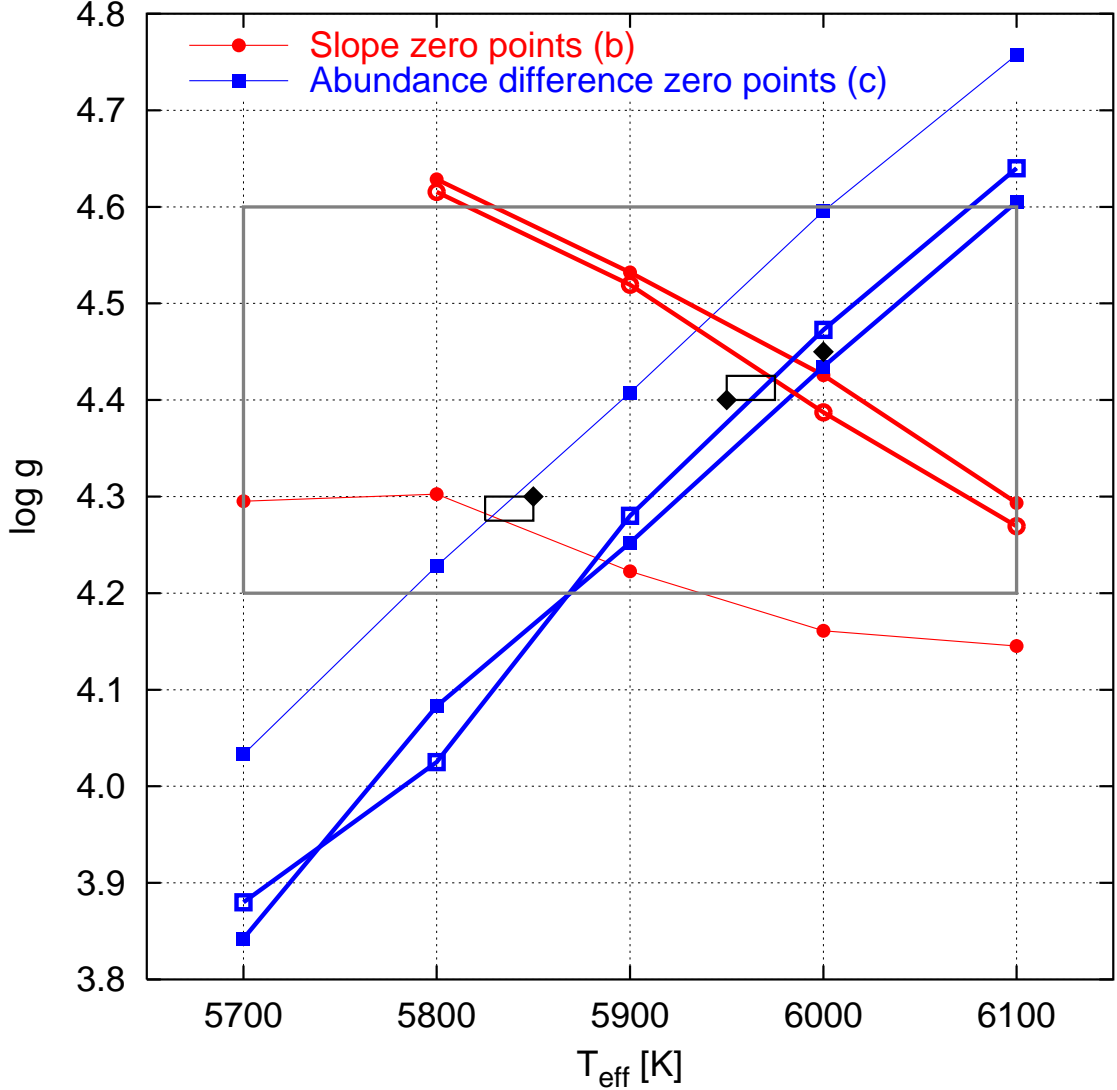} }
\caption{Determination of the best-fit atmospheric parameters for HD\,68988. Circles represent (interpolated) models for which the same value of \vmicro\ eliminates any trend between Fe~I line abundance and equivalent width and between line abundance and excitation energy (step b in Section~\ref{parameters}). Squares represent (interpolated) models for which the difference between Fe~I and Fe~II abundance is zero (step c in Section~\ref{parameters}, \vmicro~=~1.0~\kms). The intersection of two corresponding lines gives the best-fit parameters, which are rounded to the location indicated by diamonds. Thin and thick lines correspond to model metallicities of +0.0 and +0.5~dex, respectively. Thick lines with open symbols are obtained when model atmospheres calculated with the ATLAS9 code are used (see text, Section~\ref{parameters}). The gray rectangle indicates the area of the calculated grid.}
\label{analysis1}
\end{figure}

We estimate the {\em internal} errors in our final spectroscopically determined parameters to be 50~K for \Teff, 0.05~[cgs] for \logg, and 0.05~\kms\ for \vmicro.
Figure~\ref{analysis1} illustrates the last step of the procedure ({\em d}), ie. the (\Teff, \logg) sets determined from steps {\em b} and {\em c}, for the star HD\,68988.
We used a grid of $5 \times 5$ model atmospheres centered on (5900, 4.4), which is indicated by the gray rectangle.
The initial parameters were determined by method 3 (Section~\ref{init}).
Calculations using two different metallicities (ODFs) are shown connected by thin and thick lines.
The intersection of two corresponding lines gives the best-fit parameter-pair -- (5850, 4.30) for a solar ODF, with \vmicro~=~0.7~\kms\ and [Fe/H]~=~+0.35, and (6000, 4.45), \vmicro~=~1.35~\kms\ and [Fe/H]~=~+0.36 for an ODF of +0.5.

This result does not change significantly when we use model atmospheres calculated with the ATLAS9 code \citep{Kuru:93,Heit:02} instead of the \citetalias{Gust:75} code.
This is shown by the open symbols in Fig.~\ref{analysis1}, which give a solution of (5950, 4.40), \vmicro~=~1.30~\kms\ and [Fe/H]~=~+0.35.
The main differences between the two types of model atmospheres are the following:
(1) Differences in the ODFs -- number and types of transitions, wavelength resolution, constant ``doppler broadening velocity'' of 2~\kms, which includes both the thermal and non-thermal (\vmicro) contributions to the mean velocity of the atoms in the \citetalias{Gust:75} code vs. constant \vmicro\ in ATLAS9 (1~\kms\ used here),
(2) Different optical depth ($\tau$) scale -- \citetalias{Gust:75}: $-4.4 \le \log \tau \le -0.6$ in steps of 0.2 and $-0.6 \le \log \tau \le +1.2$ in steps of 0.1; ATLAS9: $-6.875 \le \log \tau \le +2.0$ in steps of 0.125,
(3) Convection treatment -- \citetalias{Gust:75}: mixing length theory (MLT) with parameters $\alpha = 1.5, \nu = 8, y = 0.076$; ATLAS9: convection model described by \citet[ CGM]{Canu:96}, which gives a temperature structure very similar to MLT with $\alpha = 0.5$ \citep[see][]{Heit:02}.
We also performed the same test using ATLAS9 with MLT and $\alpha = 1.25$ and obtained exactly the same results.
Therefore, point (3) seems to be of minor significance.

\begin{figure}
\resizebox{\hsize}{!}{ \includegraphics{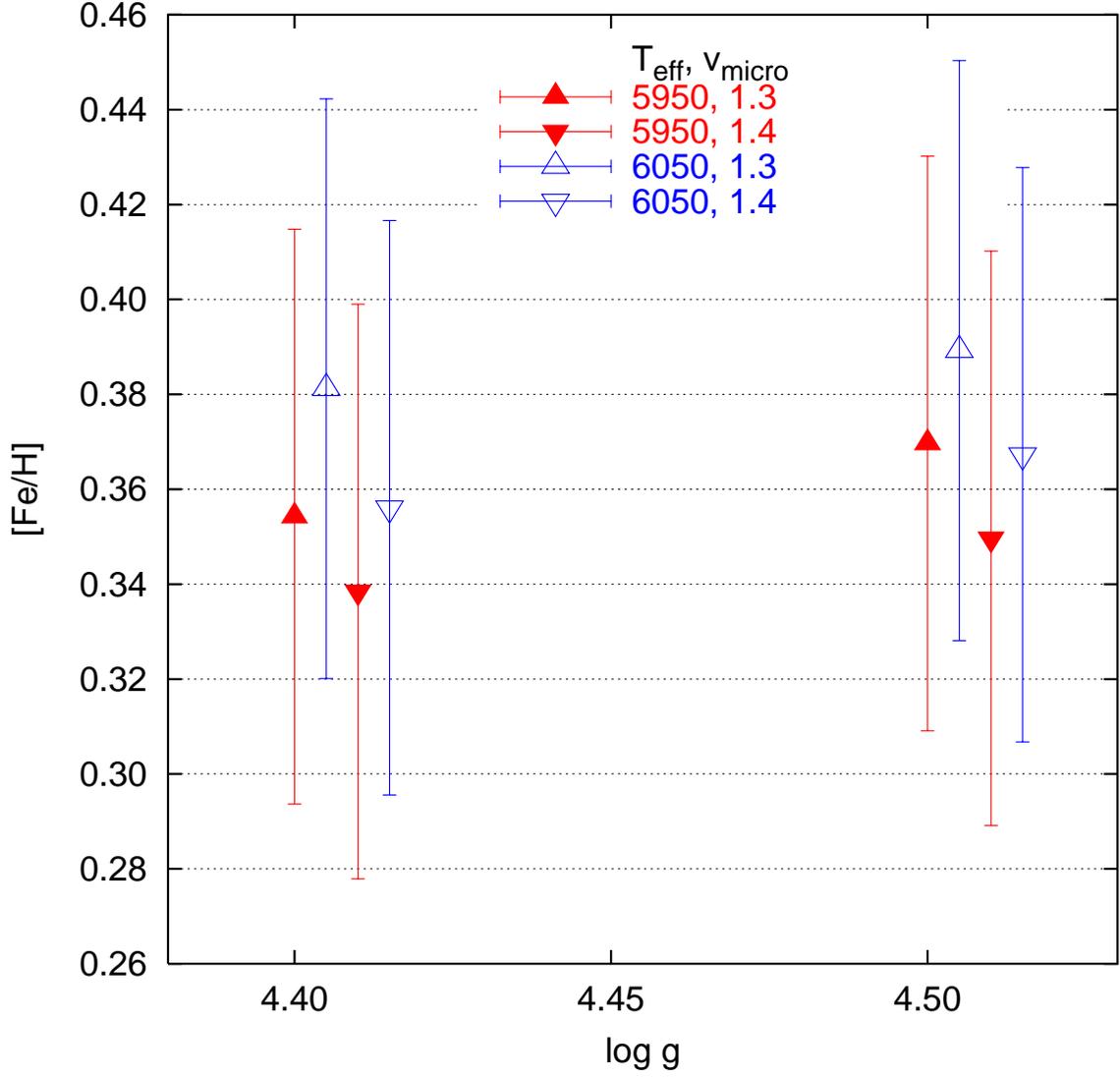} }
\caption{Variation of [Fe/H] when \Teff, \logg\ and \vmicro\ are changed by $\pm$50~K, $\pm$0.05~[cgs], and $\pm$0.05~\kms\, respectively, from the best-fit values of (6000, 4.45, 1.35) for HD\,68988. Models with equal \logg\ but different \Teff\ and \vmicro\ have been shifted in \logg\ for better visibility.}
\label{analysis2}
\end{figure}

Figure~\ref{analysis2} shows the variation of the mean iron abundance when the parameters are changed by $\pm$50~K, $\pm$0.05~[cgs], and $\pm$0.05~\kms\, from the best-fit values for HD\,68988. It shows that the abundance changes are much smaller than the scatter of the line abundances (indicated by the error bars). 
The parameters, including the iron abundance, derived for all program stars are listed in Tables~\ref{param_tab_1} to \ref{param_tab_4}.

For eight stars a value of 0.00 was used for \vmicro. In these cases trends of line abundances versus equivalent widths could not be removed with a positive value of \vmicro, and in some cases the formal solution determining \vmicro\ by extrapolation would even have resulted in negative values. This indicates a problem with modeling the equivalent widths for these stars, all of which are rather cool (\Teff $\le$ 5400~K). Varying the microturbulence parameter with depth in the atmosphere would probably alleviate this problem, but would require too many more free parameters to be determined. See also the discussion by \citet{Take:02b}.
We note that for all these cases the the exact value of \vmicro\ is not critical in determining the abundances. For example, changing \vmicro\ from 0.0 to 0.5 reduces [Fe/H] by only 0.02 or 0.03~dex for almost all stars (0.04~dex for HD\,10700). 

\subsection{Reference Model Parameters}

To examine the influence of the parameters chosen to calculate the solar line abundances on the parameter determination, we recalculated these abundances using solar models with parameters given in the first three columns of Table~\ref{reference}. These values lie within the range of parameters determined or quoted by \citet[ their Table~2]{Take:02a}. Then we repeated the analysis for the star HD\,68988 relative to these abundances. The last three columns of Table~\ref{reference} list the solutions obtained for each reference model. It shows that changing a particular parameter for the reference model mainly influences the corresponding parameter of the star under investigation. Moreover, each best-fit model leads to the same mean iron abundance of +0.37~dex. 

\subsection{Comparison with Photometric Calibrations}
\label{comparison_phot}

\begin{figure}
\resizebox{4.4in}{!}{ \includegraphics{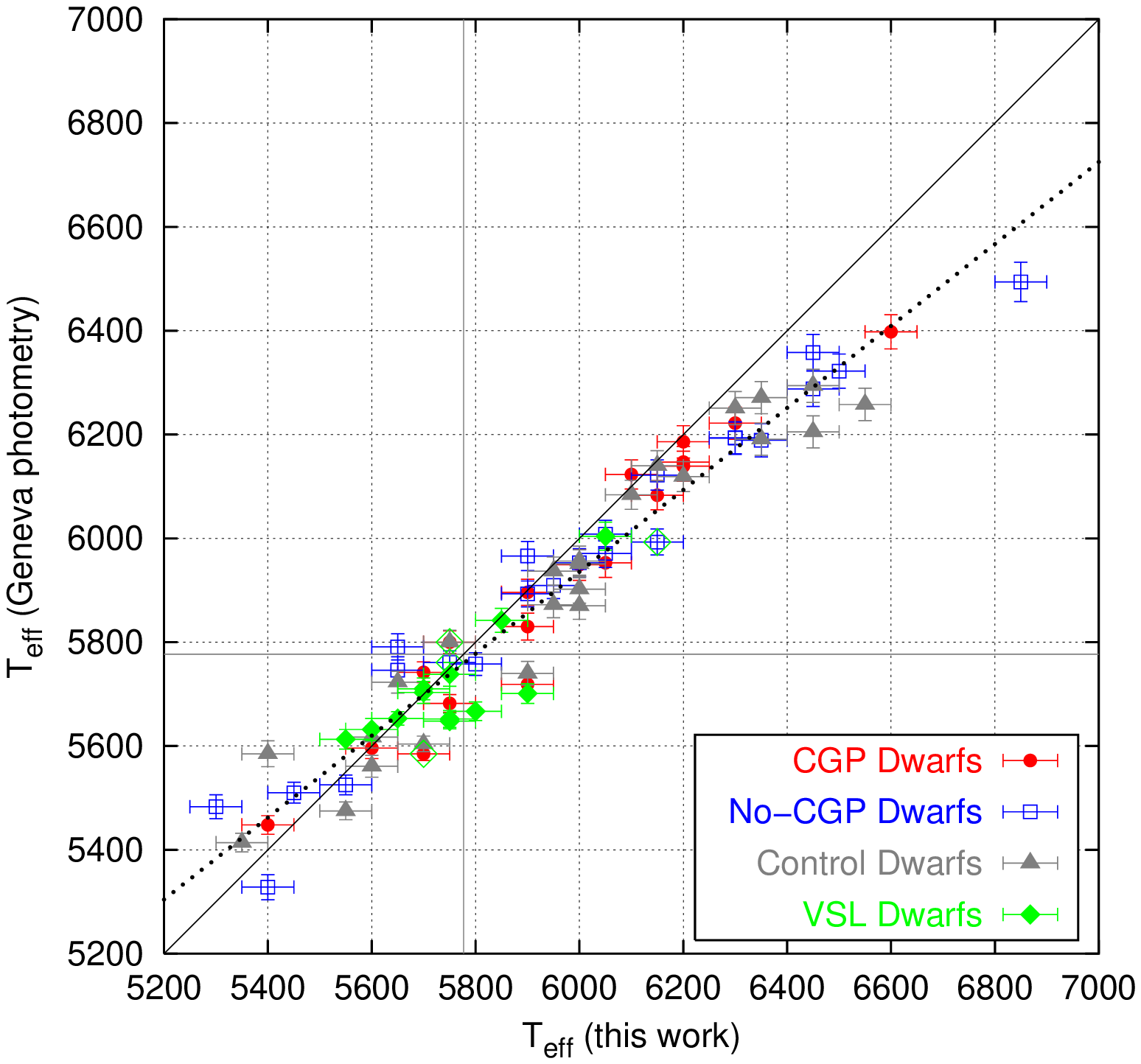} }\\
\resizebox{4.4in}{!}{ \includegraphics{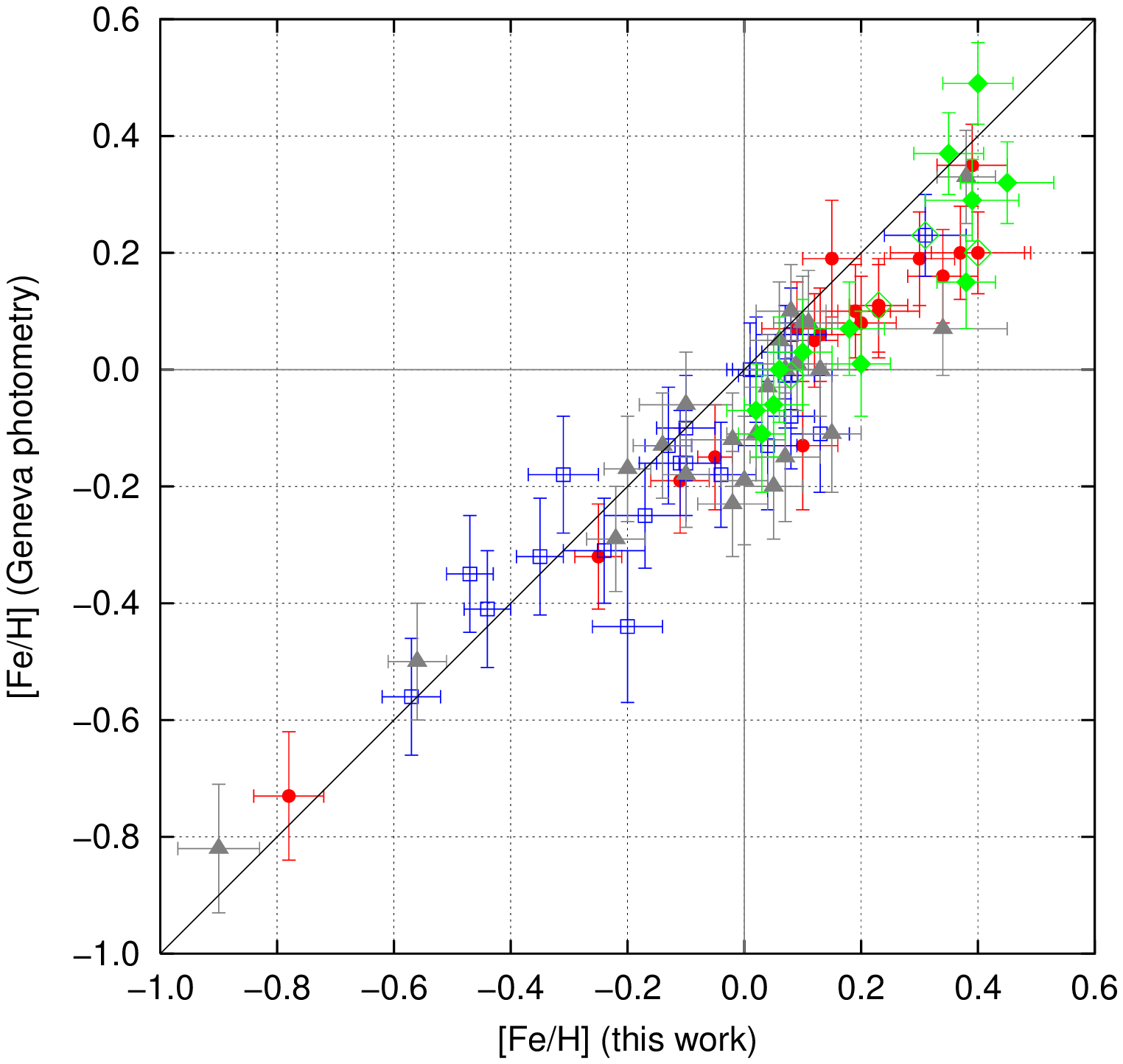} }
\caption{Comparison of our spectroscopic effective temperatures (upper panel) and iron abundances (lower panel) with that derived from Geneva photometry (see text, Section~\ref{comparison_phot}). Gray horizontal and vertical lines are placed at the solar value. The dotted line represents a linear fit to the data. CGP and no-CGP dwarfs which are also VSL dwarfs are marked with a large green open diamond.}
\label{comp_ge}
\end{figure}

Photometric measurements in the Geneva system are available for 77 of our program stars \citep{Merm:97}, and we used the calibration of \citet{Kuen:97} to calculate effective temperatures, gravities and metallicities for these stars. In Fig.~\ref{comp_ge} we compare the photometric \Teff\ values with the spectroscopically derived ones (upper panel). Averaging over all data points, the difference between the two temperature scales is small (52~$\pm$96~K), but they do not follow a one-to-one relationship with a systematic offset. Instead, as indicated by the dotted line, the one-to-one correspondence appears to be rotated around the solar value. Towards larger temperatures, the spectroscopic analysis leads to increasingly higher temperatures than the photometric calibration (linear fit: \Teff$^{\rm G}$ = 0.79 \Teff$^{\rm spec}$ + 1200). The comparison of metallicities (Fig.~\ref{comp_ge}, lower panel) displays a more complicated pattern, but essentially shows an offset of +0.08~$\pm$0.09~dex of the spectroscopic abundances with respect to the photometric ones. However, all of the eight stars with [Fe/H]$\lesssim-$0.3~dex behave in the opposite way, the photometric abundances being (slightly) larger than the spectroscopic ones. The surface gravity values are on the average very similar, but with a large scatter (+0.07~$\pm$0.39~[cgs]).

\begin{figure}
\resizebox{4.4in}{!}{ \includegraphics{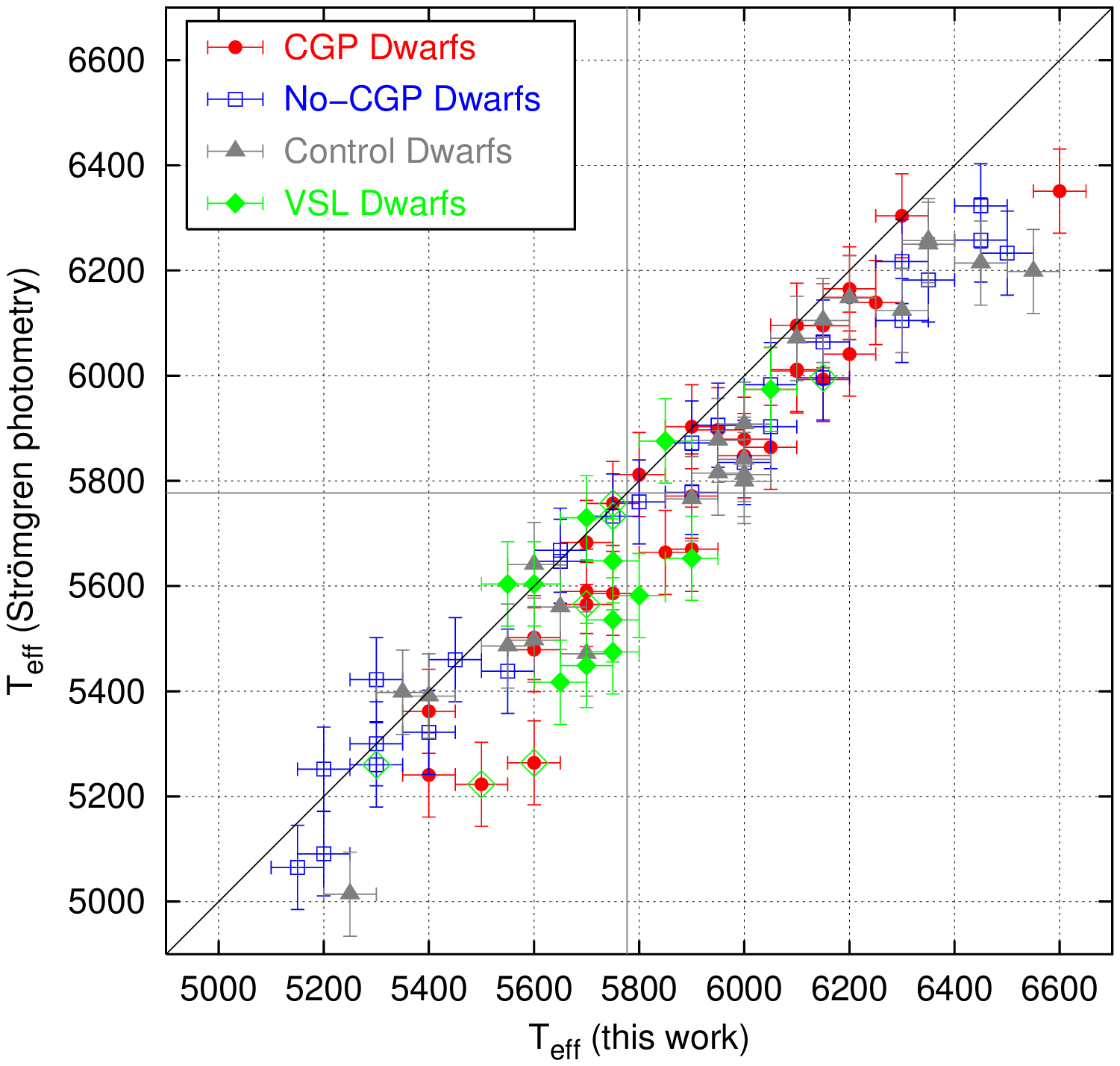} }\\
\resizebox{4.4in}{!}{ \includegraphics{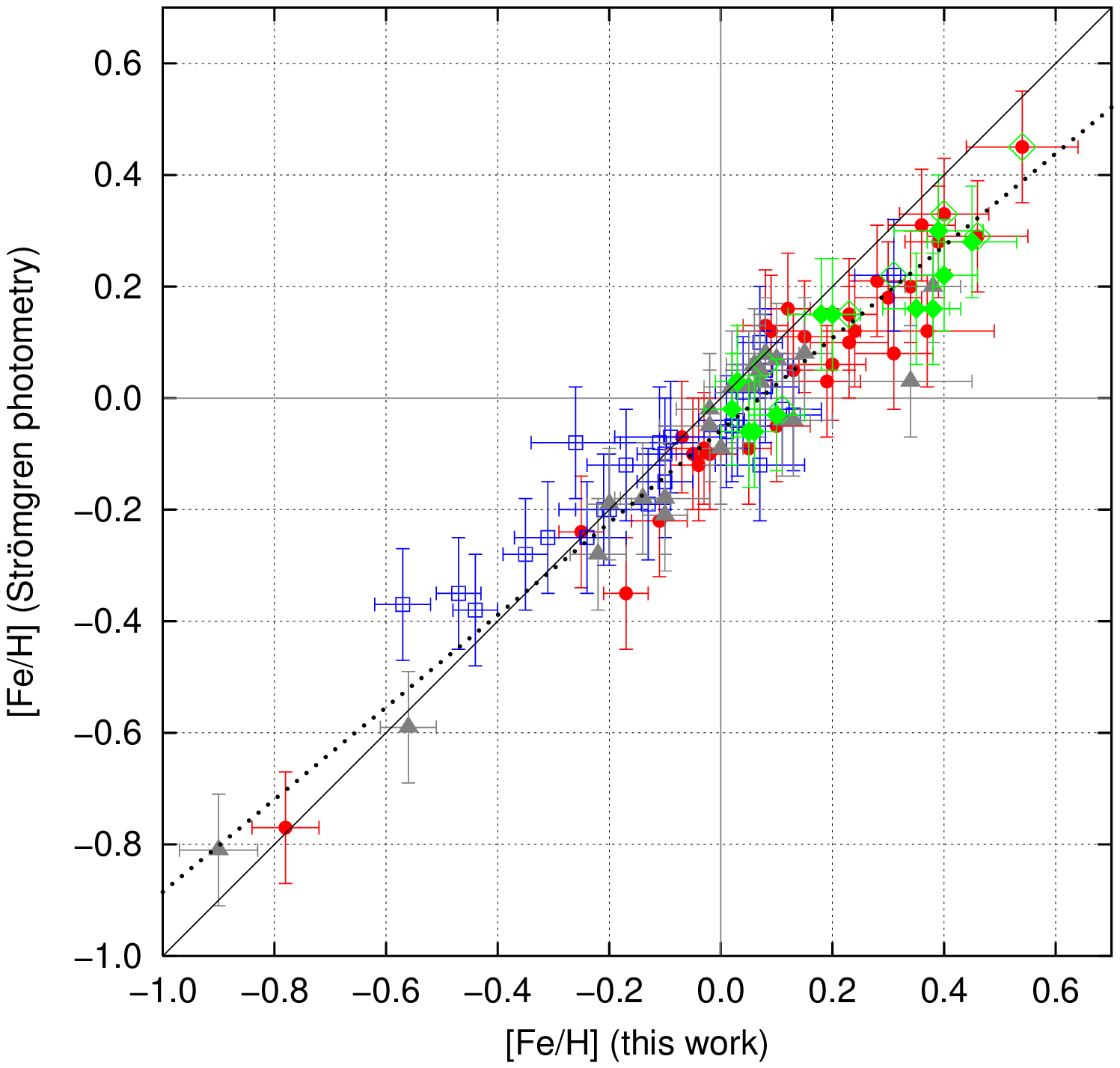} }
\caption{Comparison of our spectroscopic effective temperatures (upper panel) and iron abundances (lower panel) with that derived from Str\"omgren photometry (see text, Section~\ref{comparison_phot}).}
\label{comp_st}
\end{figure}

The most recent calibration for Str\"omgren photometry has been published by \citet{Mart:02}.
We applied it to 95 of our program stars. The data have again been taken from the GCPD \citep{Merm:97}. Fig.~\ref{comp_st} shows how the photometric \Teff\ and [Fe/H] compare to the spectroscopic ones derived in the present work. The photometrically derived temperatures are systematically lower by 105~$\pm$94~K, while the photometric metallicities are on average lower by 0.06~$\pm$0.09~dex, with larger differences for higher [Fe/H] (dotted line in Fig.~\ref{comp_st}, linear fit: [Fe/H]$^{\rm S}$ = 0.83 [Fe/H]$^{\rm spec} -$ 0.06).
These differences are somewhat surprising, since the calibration is entirely based on field stars with spectroscopically determined metallicities and has been derived in order to improve the metallicity accuracy for metal-rich stars. However, the spectroscopic parameters were taken from the compilation by \citet{Cayr:01}, using all available values (on average 2.3 per star) with equal weight, without examining this rather inhomogeneous dataset for any systematic effects. Note that \citet{Laws:03} also obtained photometric metallicities on average 0.1~dex lower than spectroscopic ones for 69 stars with planets, using in part the \citet{Mart:02} calibration.

These differences have to be considered when comparing these small samples of stars to large stellar samples which represent the solar neighborhood as completely as possible (see Section~\ref{stat}), and for which only photometric abundances are available up to now. 

\subsection{Comparison with Spectroscopic Studies}
\label{comparison}

\begin{figure}
\resizebox{4.4in}{!}{ \includegraphics{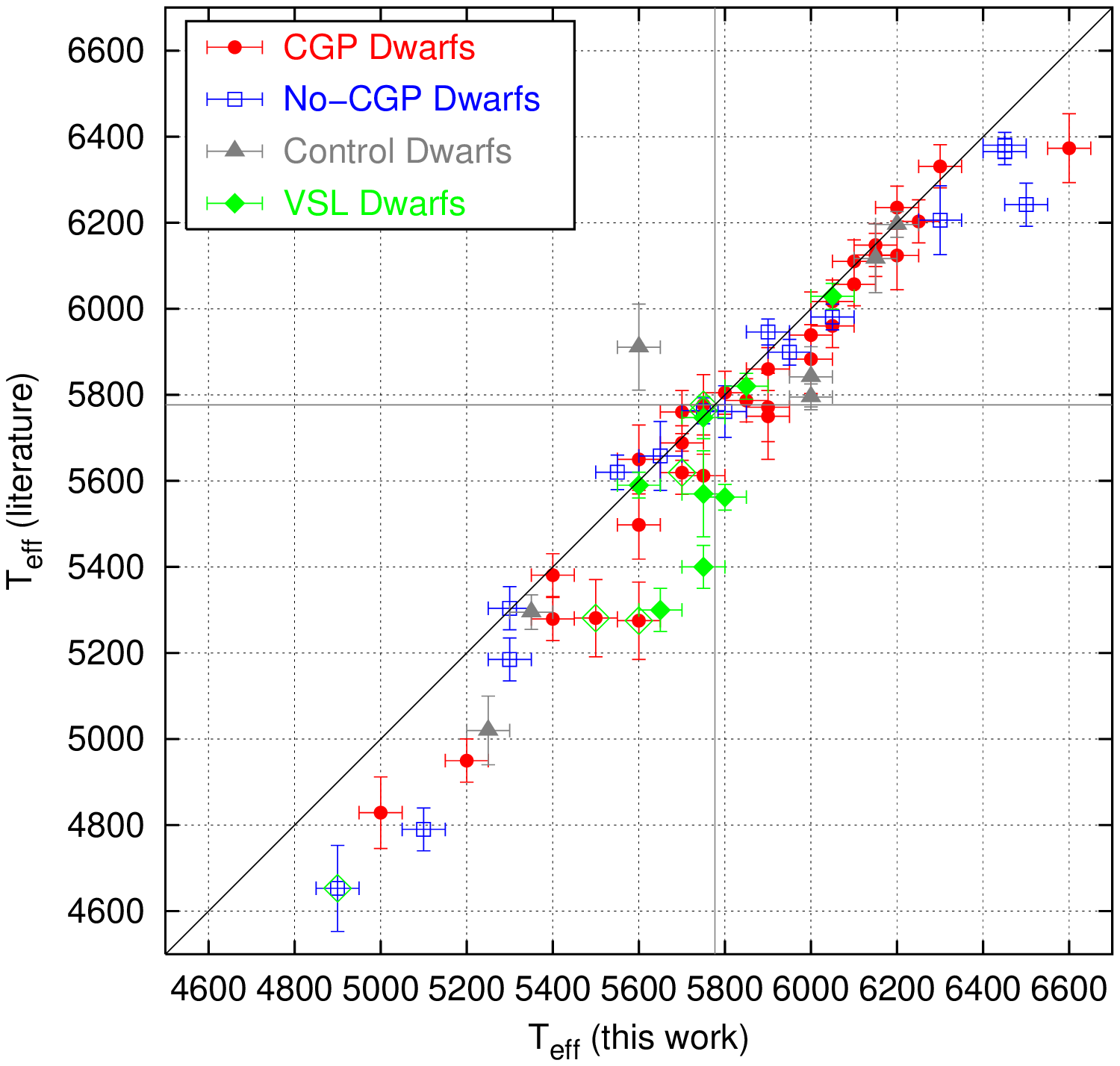} }\\
\resizebox{4.4in}{!}{ \includegraphics{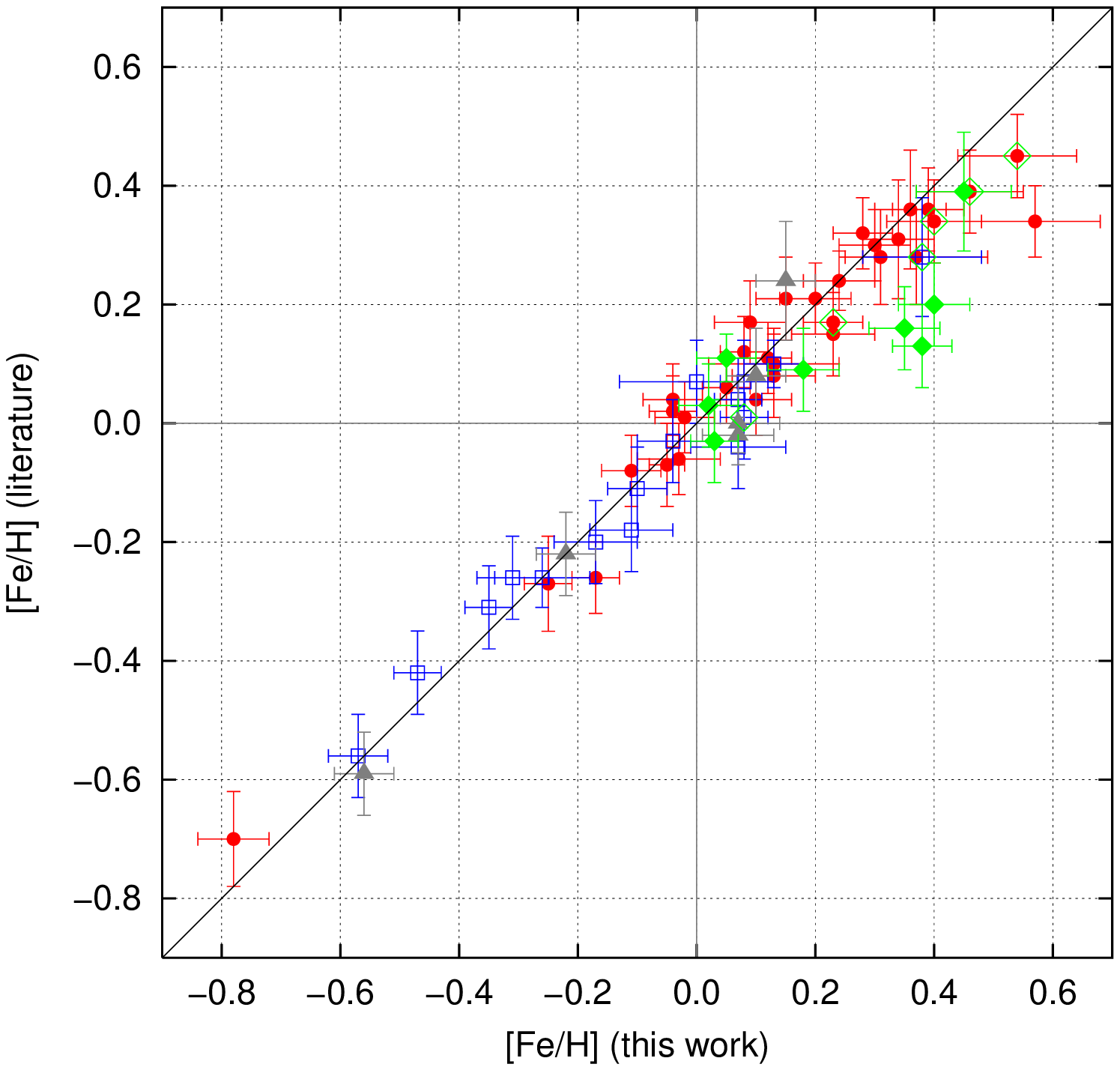} }
\caption{Comparison of our effective temperatures (upper panel) and iron abundances (lower panel) with spectroscopic literature values (see text, Section~\ref{comparison}).}
\label{comp}
\end{figure}

In Fig.~\ref{comp} we compare the derived effective temperatures and iron abundances with results from abundance analyses by \citet{Edva:93a,Fuhr:98,Felt:98}, Gonzalez et al. (1997-2001), \citet{Sant:00,Sant:01b,Sant:01a,Felt:01,Gaid:02,Sant:03,Take:02b,Sada:02}. Note that previous abundance analyses are available for almost all CGP dwarfs -- two stars have been analyzed for the first time in the present work (HD\,40979 and HD\,72659). A parameter comparison can be made for about half (16) of the no-CGP dwarfs, as well as seven control and eight VSL dwarfs. In total, data for 63 stars are shown in Fig.~\ref{comp}. When a star has been analyzed by more than one author, the values have been averaged.

For most stars, the effective temperature determinations agree reasonably well, but there are some cases in which the temperatures derived by us are up to 350~K greater than the values from the literature. Almost all of these stars lie near a border in the parameter space occupied by the total sample: they are either very cool (\Teff $\le$ 5250~K) or very metal rich ([Fe/H] $\ge$ +0.35~dex). However, the iron abundances agree very well for all stars except for three metal rich VSL dwarfs, one control dwarf and one CGP dwarf.
These are HD\,99491 \citep{Felt:01}, HD\,103095 \citep[$\lbrack$Fe/H$\rbrack$=$-$1.25, not shown in the lower panel of Fig.~\ref{comp}]{Fuhr:98,Take:02b}, HD\,104304 \citep{Felt:01}, HD\,125184 \citep{Edva:93a} and HD\,177830 \citep{Gonz:01,Sant:03}, with abundances derived in the present work being greater than in the literature by 0.20, 0.17, 0.19, 0.25 and 0.23~dex, respectively.
For HD\,99491 the difference reduces to +0.10~dex if we use the the parameters of \citet[ see Table~\ref{diff}]{Felt:01}, but we obtain a difference between Fe~I and Fe~II abundances of $-$0.15~dex.
For HD\,177830, using the published parameters does not change the derived abundance while resulting in a large Fe~I $-$ Fe~II abundance difference. However, using only the lines from the published line list reduces both abundance differences. The remaining difference does not seem to be due to a difference in measured equivalent widths (Fig.~\ref{ew_dwp}).
For the other three stars we recover the published abundances reasonably well if we use the published atmospheric parameters, but for HD\,104304 we obtain a large difference between Fe~I and Fe~II abundances. All stars show considerable trends of line abundance with excitation energy (HD\,125184 also with equivalent width) when the listed parameters are used.

\begin{figure}
\resizebox{5.4in}{!}{ \includegraphics{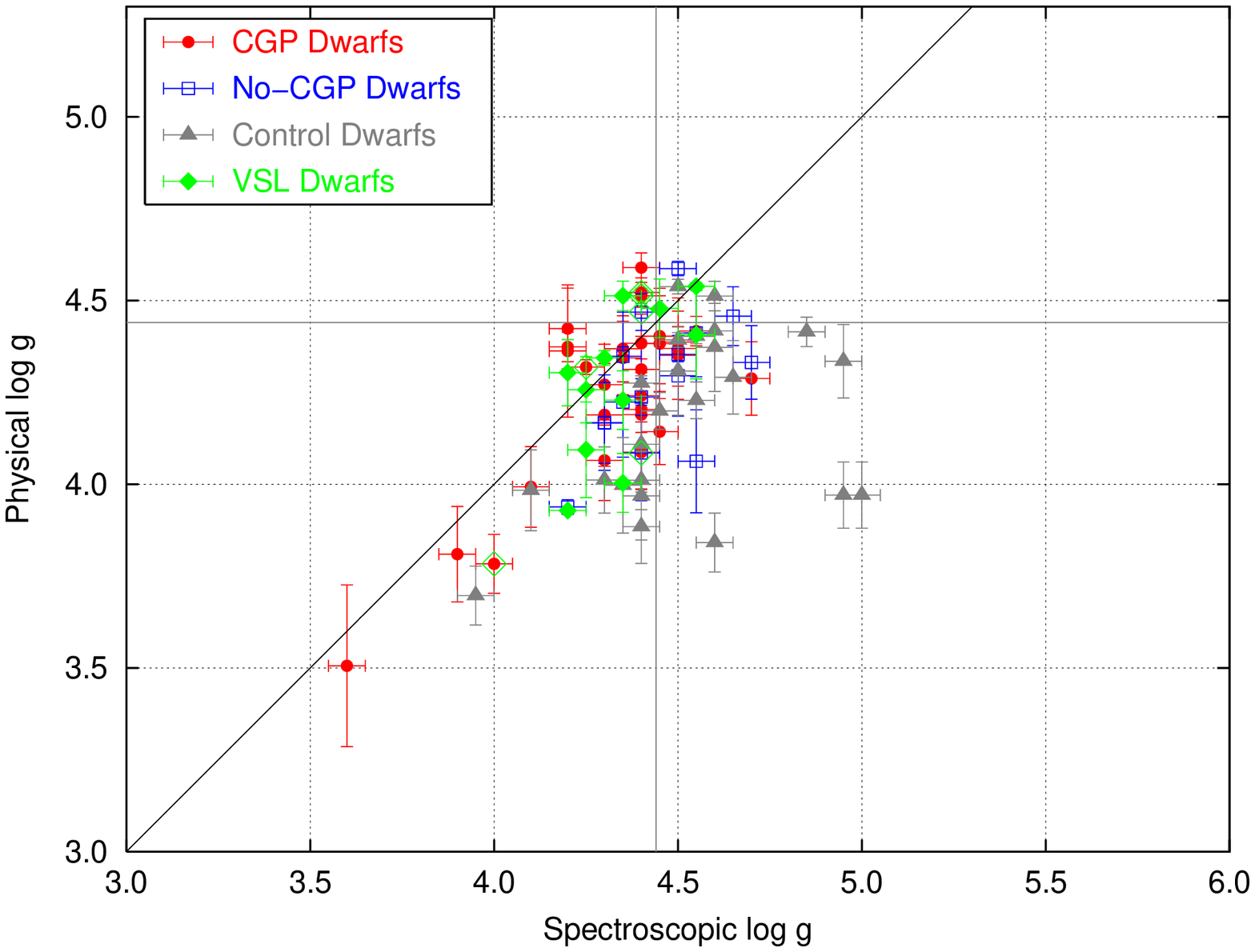} }
\resizebox{5.4in}{!}{ \includegraphics{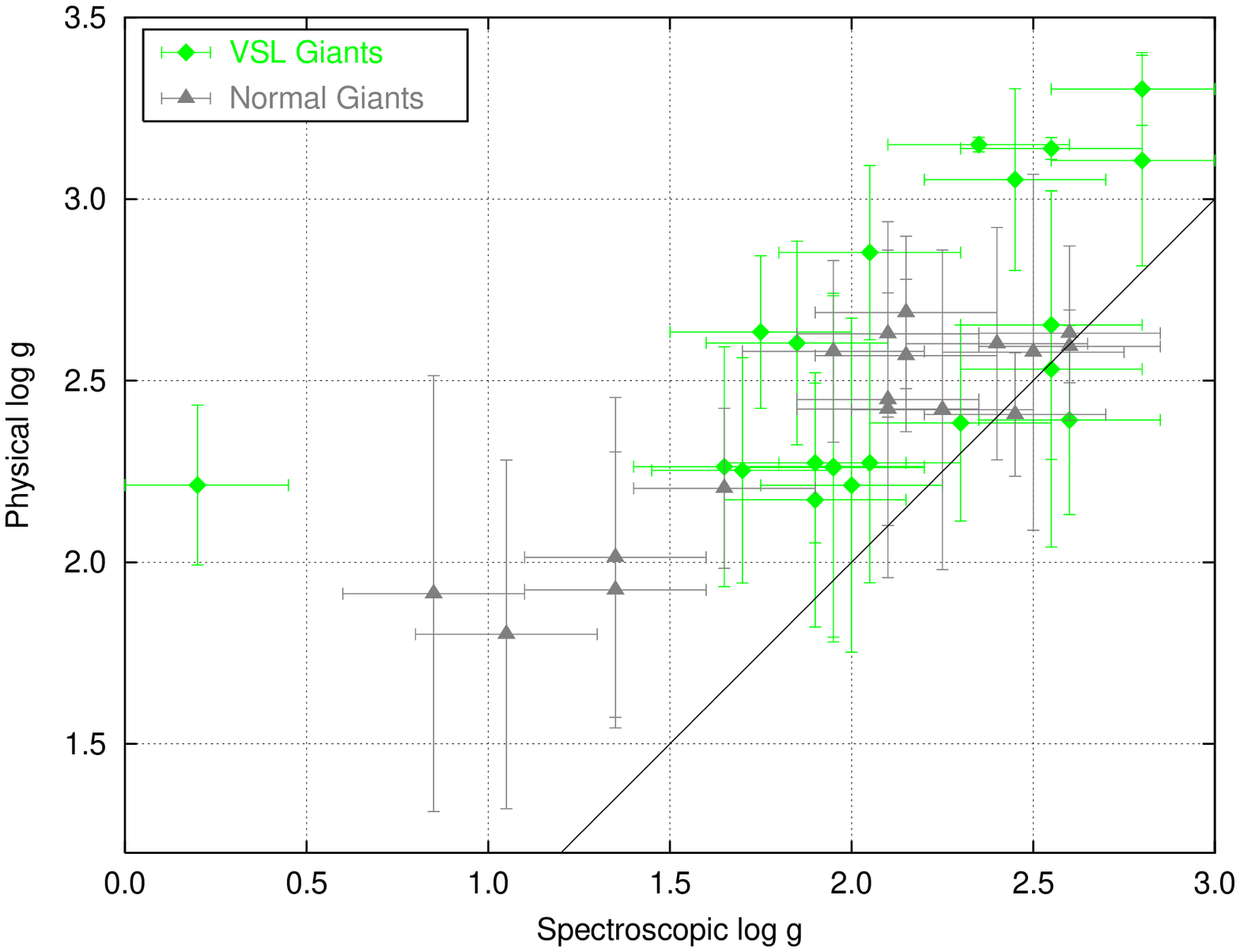} }
\caption{Comparison of \logg\ derived from high resolution spectroscopy in the present work (upper panel) and by \citet[ lower panel]{Luck:95} for giant stars with \logg\ derived from evolutionary tracks \citep{Alle:99}. Note that in both panels, both axes are on the same scale.}
\label{comp_logg}
\end{figure}

\citet{Alle:99} used B$-$V colors and parallaxes from the Hipparcos Catalogue to calculate absolute magnitudes and interpolate in the isochrones of \citet{Bert:94} in order to derive stellar parameters for stars within 100~pc from the Sun. 80 of our program stars are included in this work and we can compare our spectroscopically derived \logg\ values with these ``physical'' ones. This comparison is shown in Fig.~\ref{comp_logg} (upper panel). The physical \logg\ values have been corrected for differences in \Teff, but these corrections are very small (on average 0.01~dex).
For most stars, the two values agree within $\approx$ 0.3~[cgs], but on average the spectroscopic \logg\ values are greater than the physical ones by 0.17 $\pm$~0.23~[cgs]. There is a remarkable difference in the disagreement between the four samples: The largest differences exist for the 24 control dwarfs in common (with a mean of 0.35 $\pm$~0.27), followed by the 15 no-CGP dwarfs (0.17 $\pm$~0.15), while the difference is only 0.07 $\pm$~0.16 for the 30 CGP dwarfs as well as the 11 VSL dwarfs.

A similar comparison has been made by \citet[ cf. their Fig. 4]{Luck:95} for VSL and normal giant stars. In the lower panel of Fig.~\ref{comp_logg} we show their spectroscopic \logg\ values against values from \citet{Alle:99}. The discrepancy is much larger for these 38 stars and has opposite sign ($-0.47\pm$~0.41). On the other hand, both spectroscopic and physical gravity determinations are less accurate for the giants than for the dwarf stars.
\citet{Luck:95} quote as the main reasons for differences between spectroscopic and physical gravities (1) the accuracy of the absolute magnitudes, (2) the accuracy of the Fe oscillator strengths, and (3) the atmospheric models. Points (1) and (2) are least relevant in the present work since precise Hipparcos parallaxes have been used by \citet{Alle:99} to derive the absolute magnitudes and we used line-by-line abundance differences for the abundance analysis.
Hence the reasons for the gravity differences in the dwarf samples are likely to be found in the employed models, not only the atmospheric but also the evolutionary ones. The best solution seems to be to view \logg\ as a parameter with different meanings in the two contexts: In the atmosphere, it is used to define the pressure structure in such a way that lines from neutral and ionized atoms give the same abundances, whereas in the stellar structure it is determined by the radius which corresponds to a given temperature and luminosity.
Fortunately, the derived abundances are less dependent on \logg\ than on the other parameters: For example, extrapolating from the test case shown in Fig.\ref{analysis2}, a 0.2~dex change in \logg\ would result in an abundance change of only 0.02~dex.

\section{Discussion}
\label{discussion}

\subsection{Statistical Description}
\label{stat}

\begin{figure}
\resizebox{\hsize}{!}{ \includegraphics{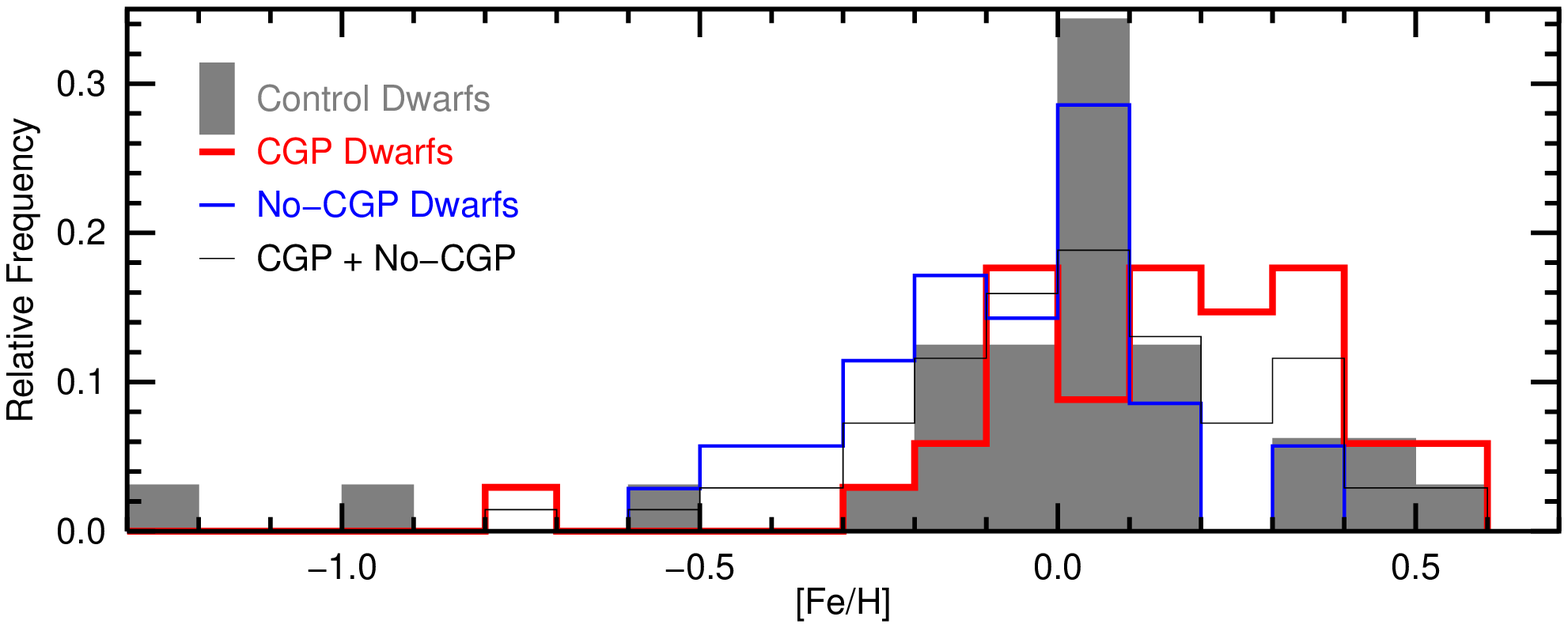} }
\resizebox{\hsize}{!}{ \includegraphics{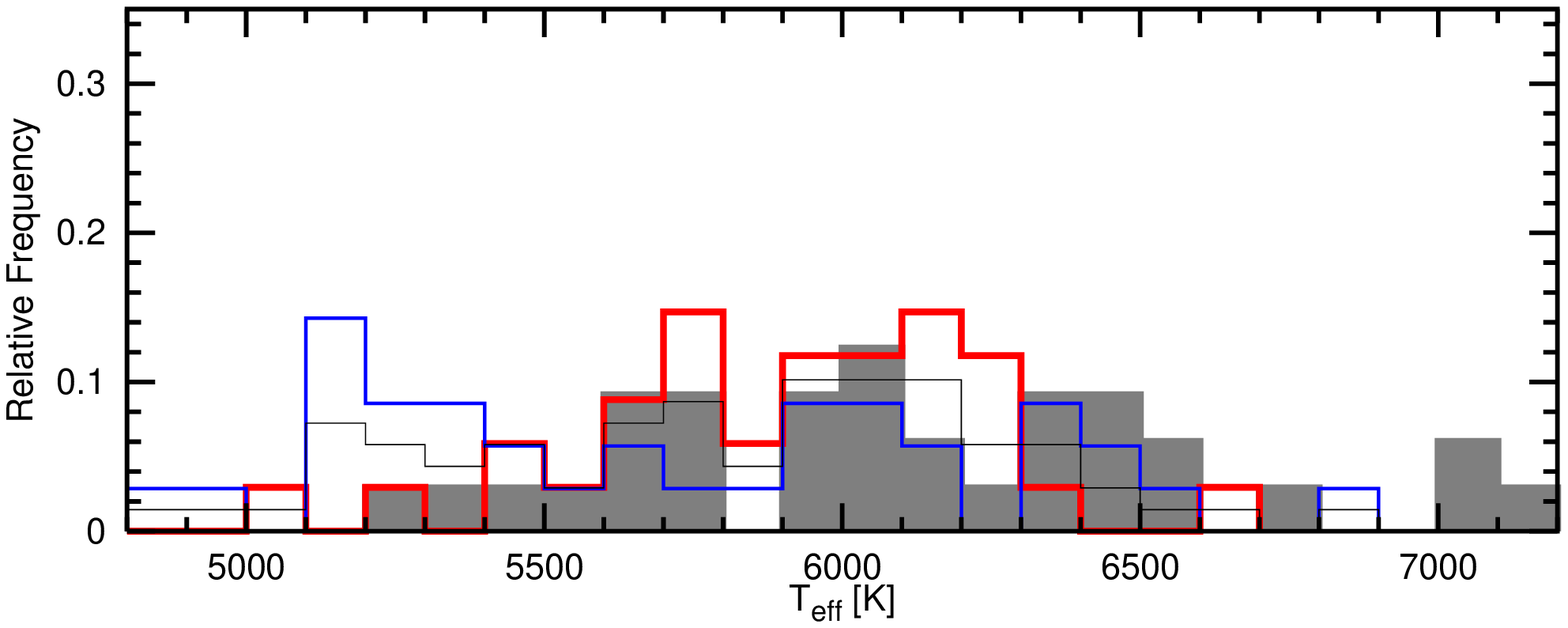} }
\caption{Abundance and temperature distributions of the program stars. The line labeled ``CGP + No-CGP'' shows the distribution for the combined data of CGP and no-CGP dwarfs.}
\label{histograms}
\end{figure}

\begin{figure}
\resizebox{4.4in}{!}{ \includegraphics{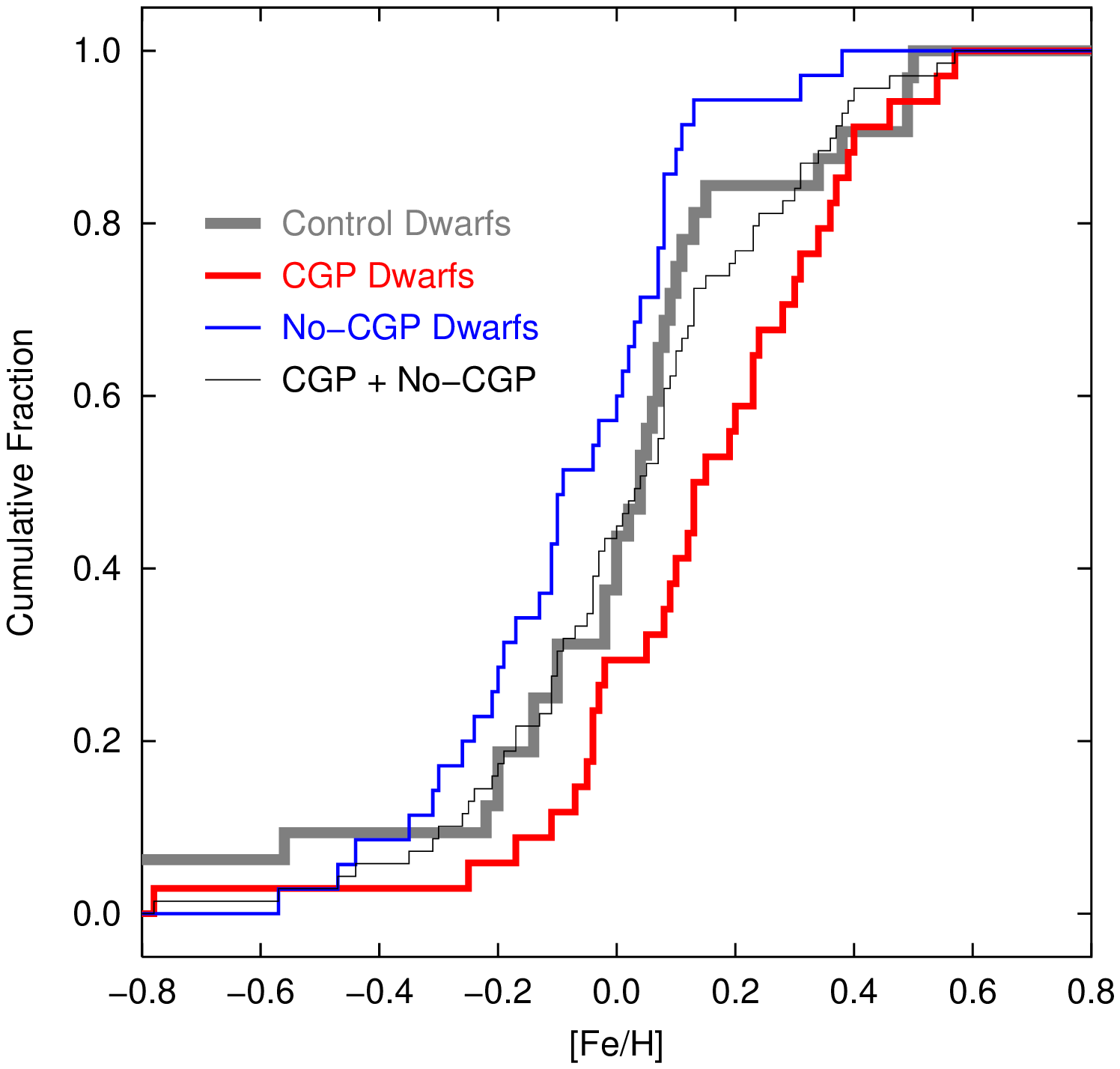} }\\
\resizebox{4.4in}{!}{ \includegraphics{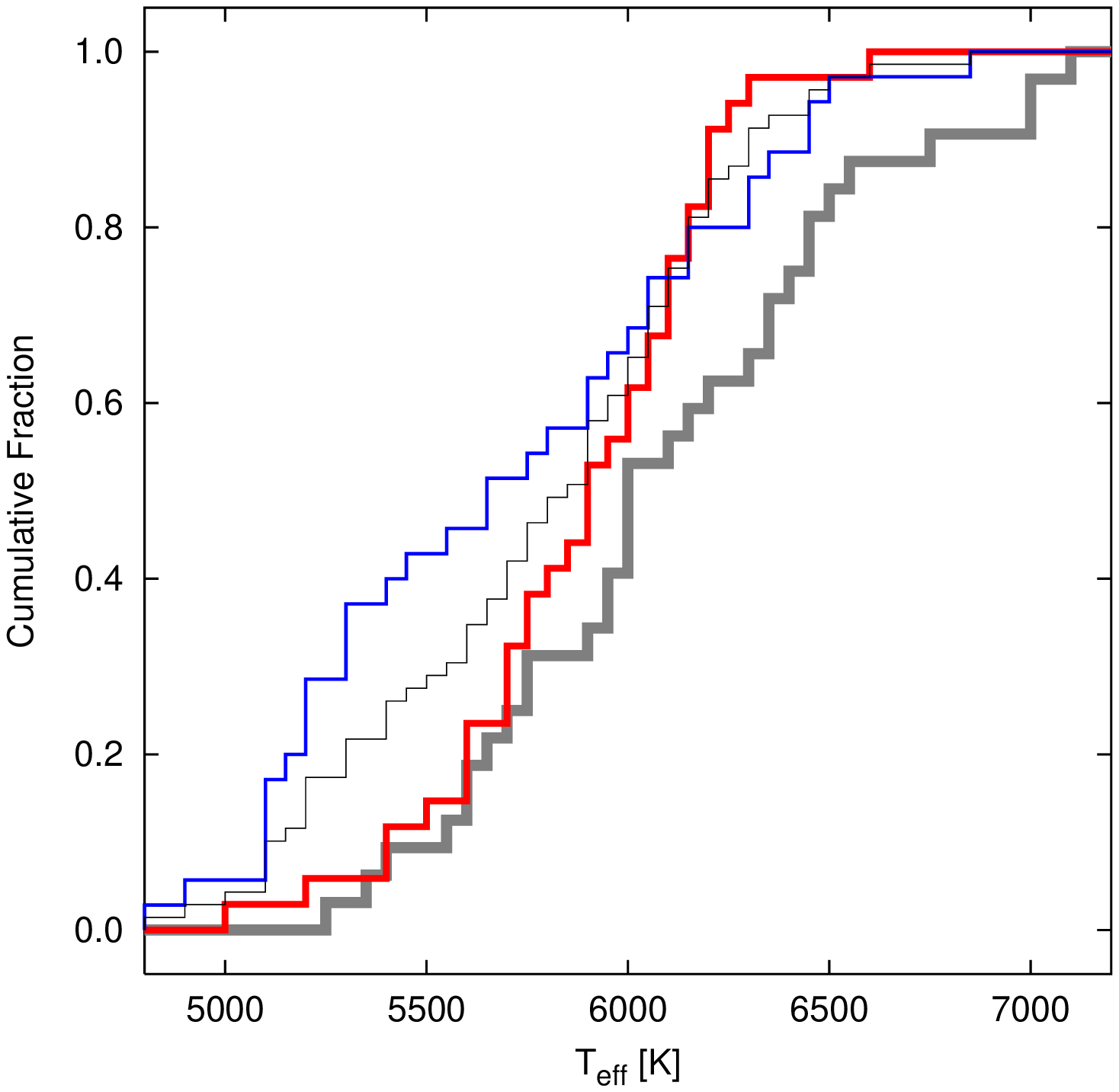} }
\caption{Cumulative distribution functions for abundances and temperatures of the program stars. Lines are as in Fig.~\ref{histograms}.}
\label{cdfs}
\end{figure}

Table~\ref{stat_tab} lists the mean iron abundances and effective temperatures for the stars in each group as well as the combination of CGP and no-CGP dwarfs. 
Figure~\ref{histograms} shows a graphic representation of the abundance and temperature distributions for these groups (except for the VSL dwarfs).
First of all we note that the mean iron abundance of the control dwarf sample is solar, although with a considerable spread of about 0.4~dex.
The results confirm \citep[e.g.][]{Sant:03} that the iron abundances of CGP dwarfs are on average greater than that of no-CGP dwarfs, namely by 0.22~dex. On the other hand, the distributions of these two groups are rather broad: a comparable number of stars from each group have abundances within $\pm$0.2~dex from solar (17 CGP and 23 no-CGP dwarfs). However, at the low and high ends of the abundance distribution the differences are remarkable: there are two CGP and ten no-CGP dwarfs with [Fe/H] $\le -0.2$~dex, but 15 CGP and two no-CGP dwarfs with [Fe/H] $\ge +0.2$~dex. The cumulative distribution functions (Fig.~\ref{cdfs}, upper panel) and Kolmogorov-Smirnov (KS) tests confirm that the [Fe/H] distributions of these two groups are not the same (the probability is only $10^{-4}$).
The standard deviations of CGP and no-CGP dwarfs are somewhat smaller than that of the control dwarfs (about 2/3) and both of the former individually are not equally distributed as the control dwarfs, although with different KS probabilities (1 and 18~\%, respectively). However, when we combine the iron abundances of the two samples, we obtain a distribution that is very similar to that of the control dwarfs (thin black line in Figs.~\ref{histograms} and \ref{cdfs}; probability = 82~\%).

We now compare the effective temperature distributions of the three samples. Although the mean temperatures of the three samples are not very different (increasing by $\approx$200~K each from no-CGP to CGP to control dwarfs) and have comparable standard deviations (see Table~\ref{stat_tab}), a KS test shows that all three distributions are quite different (null hypothesis probability between 2 and 5~\%, see also Fig.~\ref{cdfs}, upper panel). Furthermore, combining the CGP and no-CGP dwarfs gives a temperature distribution which has a low probability (8~\%) of being similar to that of the control dwarfs. Figures~\ref{histograms} and \ref{cdfs} (lower panels) show that the most significant feature leading to the differences in the distributions is the high number of no-CGP dwarfs with low temperatures: there are 16 no-CGP dwarfs with \Teff $<$ 5600~K compared to five CGP and four control dwarfs. The large fraction of no-CGP dwarfs at low temperatures seems to be complemented by a shortage of these stars in the 400~K interval centered on the solar temperature (eight, 16 and 13 stars in the same order as above). At \Teff $>$ 6000~K, there are similar numbers of stars in the three groups (eleven, 13, 15).
It thus seems that the control dwarf sample does not span the whole spectral type range of the planet search sample even though
it was selected from the same spectral type domain and
shows the same iron abundance distribution.
An explanation for this discrepancy could be the fact that the radial velocity method for planet detection reaches a maximum in accuracy for K type dwarfs \citep[see, e.g.,][ Fig. 4]{Endl:02}, whereas these stars are underrepresented in the Bright Star Catalogue due to its magnitude limit. Note that no M type stars are included in our samples.
We plan to extend the control dwarf selection for future work.
Since temperature is correlated with mass on the main sequence, the results also mean that low mass stars seem to be less likely to host close-in giant planets than solar-mass stars.
This supports the conclusion reached by \citet{Laws:03}, who studied the frequency of stars with planets among stars observed by the California and Carnegie Planet Search Program as a function of mass. They found a significantly lower fraction of stars with detected planets below 1.0~\Msun\ than between 1.0 and 1.1~\Msun.

\begin{figure}
\resizebox{\hsize}{!}{ \includegraphics{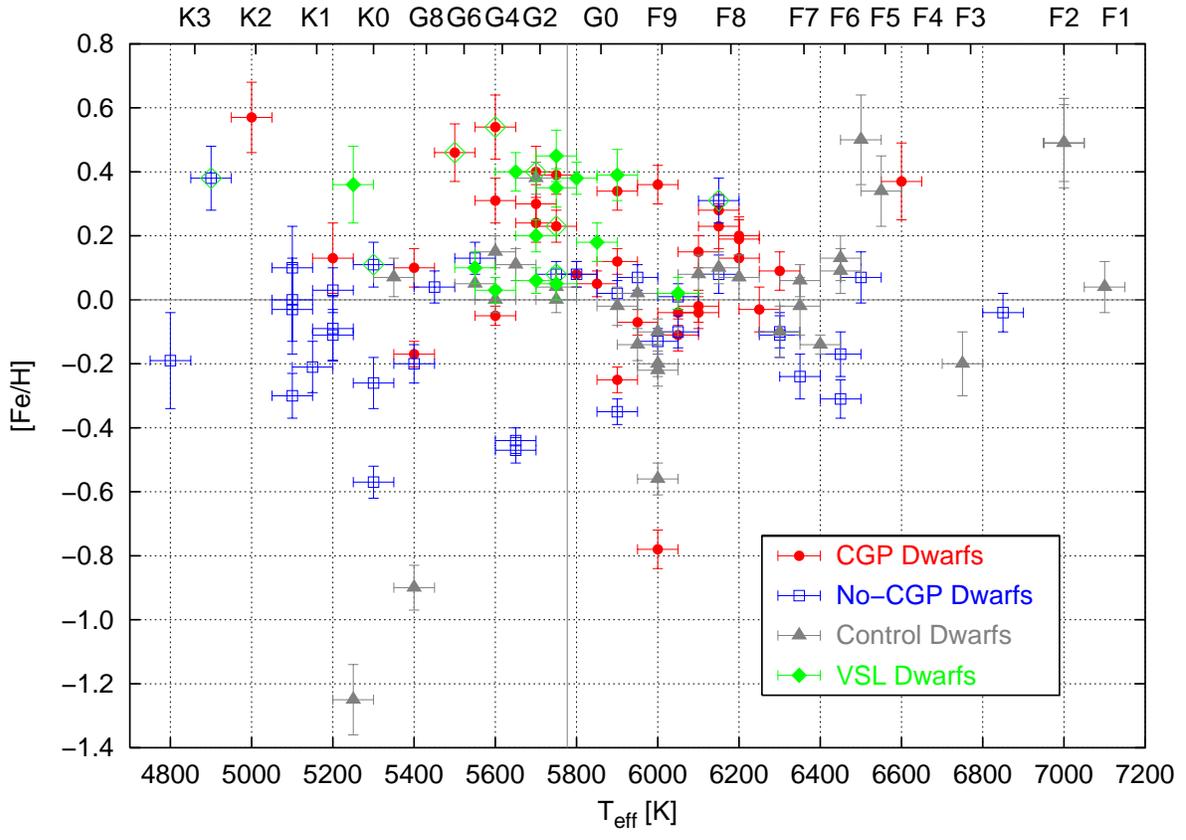} }
\caption{Differential iron abundances as a function of effective temperature. A gray vertical line is placed at the solar temperature. On the top axis, the spectral types corresponding to the temperatures according to \citet{Gray:94} are indicated.}
\label{results}
\end{figure}

\begin{figure}
\resizebox{\hsize}{!}{ \includegraphics{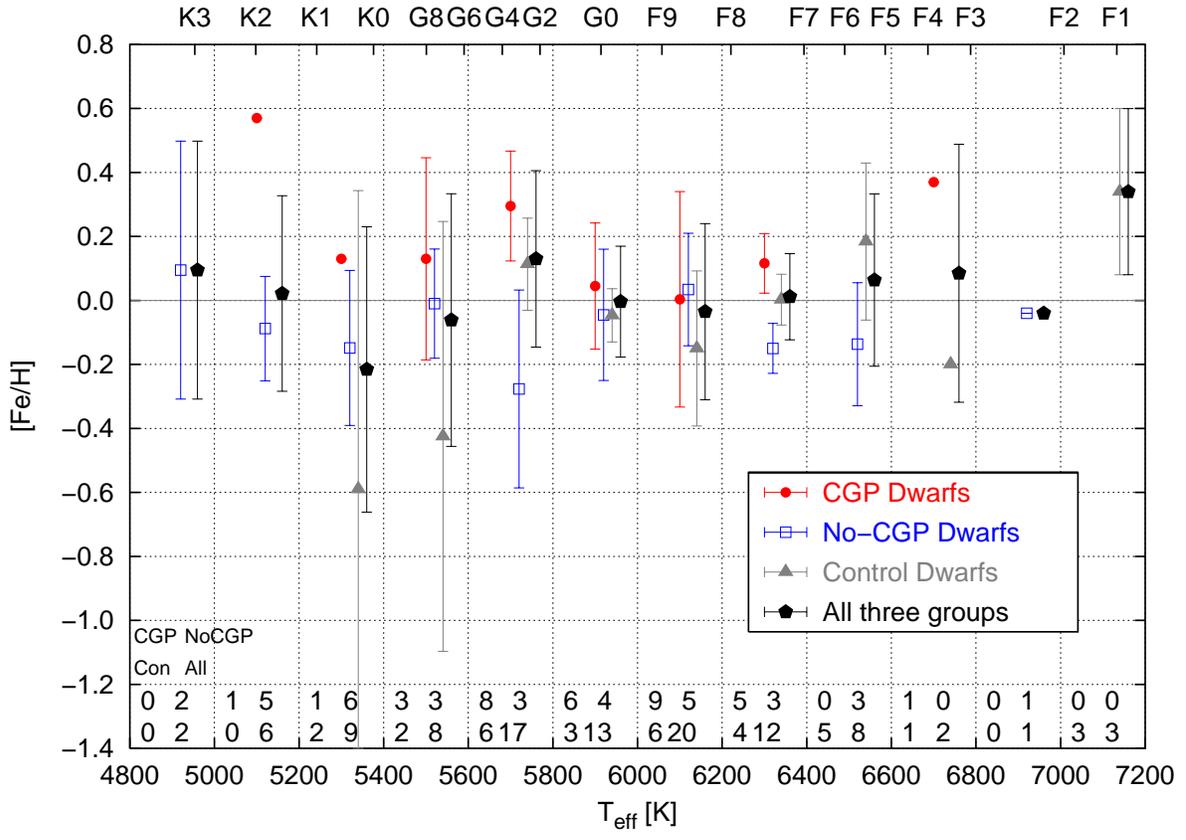} }
\caption{Mean iron abundances and standard deviations (error bars) averaged over bins of 200~K as a function of effective temperature. The numbers of stars in each bin are given at the bottom. Points for different groups have been shifted in temperature for better visibility.}
\label{results_bin}
\end{figure}

When we finally look at the abundances vs. \Teff\ for each star (Fig.~\ref{results}), we do not see any correlation between metallicity and effective temperature for any of the three groups. To see the dependence of the abundances on \Teff\ more clearly, we calculated the mean abundances and standard deviations of stars contained in bins of 200~K (shown in Fig.~\ref{results_bin}). This confirms the absence of a correlation between abundance and \Teff\ if all program stars are combined (except ``pure'' VSL dwarfs, black pentagons in Fig.~\ref{results_bin}), as well as the abundances being consistent with solar metallicity. The same can be said if control dwarfs (gray triangles) and no-CGP dwarfs (blue squares) are regarded individually, whereas the enhanced metallicity of CGP dwarfs (red circles) is clearly seen, albeit in only two out of five \Teff\ bins containing at least three stars in this group (5400~K $<$ \Teff\ $<$ 6400~K). Thus it appears that metal enrichment occurs only for CGP dwarfs with temperatures just below solar and $\approx$ 300~K higher than solar, whereas there is no difference at \Teff\ around 6000~K.

This conclusion is not changed if we exclude stars with companions with $M\sin i \ge 6.5~M_{\rm J}$. These stars might host brown dwarf candidates if the orbit inclination is, e.g., 30\degrees\ (which would result in companion masses $M \ge 13~M_{\rm J}$). 
Additionally excluding stars where the companion might be a brown dwarf candidate or M dwarf according to \citet[ based on Hipparcos astrometric data]{Han:01} does not significantly change the results either. The mentioned stars are identified in Table~\ref{param_tab_1} and the corresponding statistical characteristics are listed in Table~\ref{stat_tab}.
Because of the small numbers of stars it is not possible to draw any conclusion for cooler (\Teff\ $<$ 5400~K) or hotter (\Teff\ $>$ 6400~K) stars.

\begin{figure}
\resizebox{\hsize}{!}{ \includegraphics{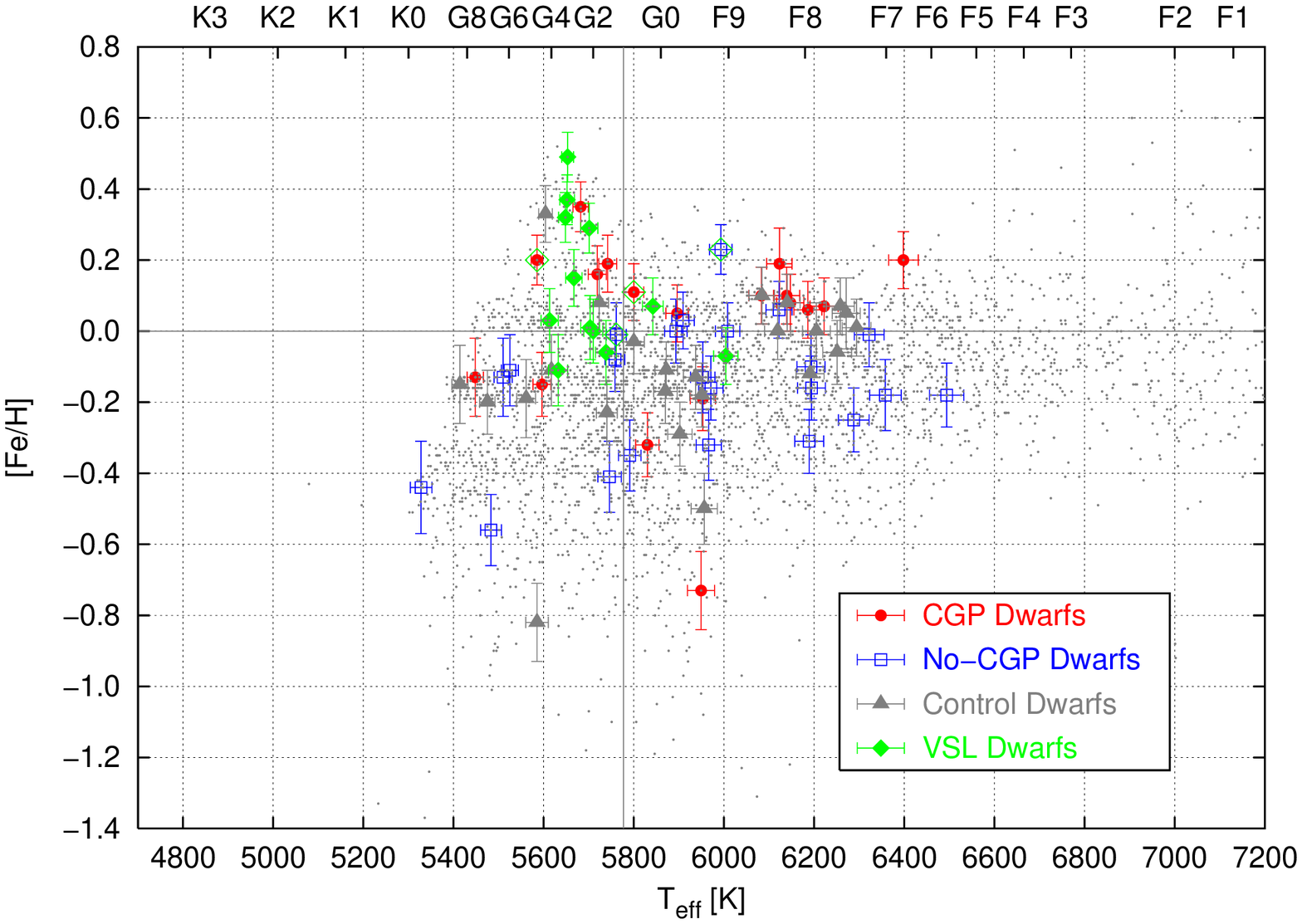} }
\caption{Iron abundances versus effective temperature, both derived from a calibration using Geneva photometry (see text, Section~\ref{comparison_phot}). Gray dots represent all stars from the Hipparcos Catalogue within a distance of 75~pc with measured Geneva photometry.}
\label{results_ge}
\end{figure}

In order to place the properties of our program stars into the larger context of solar neighborhood stars, we selected all A--K dwarfs (with $0 < M_{\rm V} < 10$) from the Hipparcos Catalogue within a distance of 75~pc and derived effective temperatures and metallicities from Geneva photometry, where available ($\approx$4160 stars), in the same way as for the planet host samples (see Section~\ref{comparison_phot}). 
Fig.~\ref{results_ge} shows the resulting \Teff -- [Fe/H] plot, on the same scale as Fig.~\ref{results}.
The left hand envelope of the data points corresponds to the borders of the domain for which the photometric calibration is valid.
Otherwise, the two most remarkable features of the plot are: (1) starting at \Teff $\approx$5600~K, the upper envelope of the 75~pc stars as well as our program stars decreases from [Fe/H] $\approx$+0.5~dex to lower metallicities when going to higher temperatures; (2) none of the no-CGP dwarfs below solar temperature have above solar metallicities.
Furthermore, one can see that below $\approx$5800~K, the photometric metallicities seem to ``cluster'' around certain values: $-$0.35, $-$0.15 and +0.05~dex, leaving ``voids'' in between. This could well be an artifact of the calibration, because this area lies at the border of its validity. Also, the [Fe/H] values of the standard stars used by \citet{Kuen:97} for the metallicity calibration ($\approx$80\% are from \citet{Edva:93a}) are not distributed normally. They also show a peak at +0.05~dex, but none at the other two values.
A comparison of Fig.~\ref{results_ge} to Fig.~\ref{results} shows the effect of the differences between spectroscopic and photometric \Teff\ and [Fe/H] (see Section~\ref{comparison_phot}). Note that the data points for our program stars seem to move closer to the solar \Teff\ and [Fe/H] when going from spectroscopic to photometric values.

\begin{figure}
\resizebox{\hsize}{!}{ \includegraphics{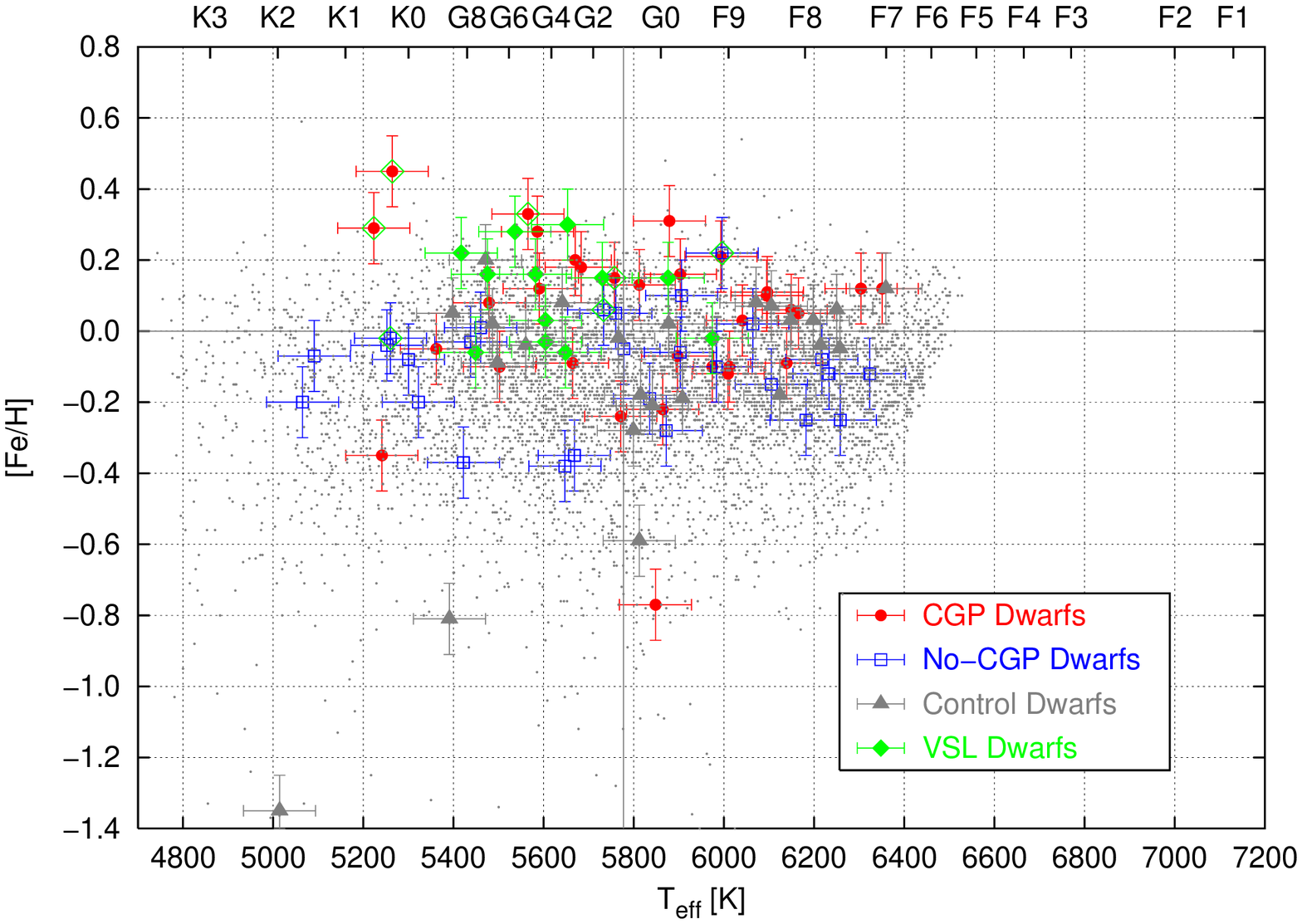} }
\caption{Iron abundances versus effective temperature, both derived from a calibration using Str\"omgren photometry (see text, Section~\ref{comparison_phot}). Gray dots represent all stars from the Hipparcos Catalogue within a distance of 75~pc with measured Str\"omgren photometry.}
\label{results_st}
\end{figure}

The same approach was taken using Str\"omgren photometry, which is available for 6130 stars within 75~pc. A look at Fig.~\ref{results_st} supports the conclusion that the structures seen for Geneva photometry are an artifact of the calibration, as the Str\"omgren calibrated values for the 75~pc stars are distributed quite uniformly. There is also no obvious trend between temperature and metallicity for the 75~pc sample. However, the increase of the maximum [Fe/H] towards cooler temperatures for the CGP dwarfs is present, as well as the lack of no-CGP dwarfs below solar temperature and above solar metallicity. Also apparent is the limit at the high temperature side, as well as the systematic difference to Fig.~\ref{results} for our program stars (see Section~\ref{comparison_phot}).

These comparisons show that the study of our program stars using photometric calibrations tend to lead to similar general conclusions, but the detailed differences between the samples vary between the three methods.

In Fig.~\ref{results} we also present the results for the stars classified as VSL dwarfs by \citet{Egge:78}.
Indeed, about 3/4 of these stars have [Fe/H] $\ge$ +0.1 (none of them have negative Fe abundances, only two have [Fe/H] $<$ +0.1 when taking into account the error bars).
When adopting the lower limit for a star to be SMR by \citet{Tayl:96} ([Fe/H] $>$ +0.2), the classification of twelve of the VSL stars can be explained by supermetallicity.
Three of these have been analyzed for the first time (HR\,582, HR\,3430 and HR\,4414\,B, in addition to four of the lower metallicity stars).
We thus show that high resolution spectroscopy is necessary to establish the nature of VSL dwarfs.
We also confirm that a considerable number of SMR dwarfs exists.
They are on average more metal rich than VSL giants \citep[$\lbrack$Fe/H$\rbrack$ $\approx$ +0.1 to +0.15~dex,][]{Luck:95}.

\subsection{Stellar Metallicity and Giant Planet Formation}
\label{formation}

How do the properties of giant planet hosts fit into the picture of giant planet formation? Two mechanisms have been proposed for giant planet formation.
The first is core accretion, where collisions of small solid particles in a protoplanetary disk form solid planet cores, which further accrete dust grains and gas until a critical mass is reached, at which point gas accretion dominates \citep[e.g.][]{Poll:96,Wuch:00}.
The second is the disk instability model, where gravitational instabilities in the protoplanetary disk lead to its fragmentation and dust grains sediment to form solid cores within the fragments \citep[e.g.][]{Boss:01,Maye:02}.
Both of these mechanisms were originally studied to explain the properties of the giant planets in our solar system and calculations were restricted accordingly to certain parameter sets.
In particular, they assumed planet formation taking place at distances from the star similar to Jupiter's distance from the Sun, and solar metallicity for the protoplanetary disk.
Recent works have explored deviations from either one or the other of these assumptions.

Within the core accretion model, \citet{Bode:00} studied if the planetary companions which have been found at distances of 0.05 to 2.1~AU from three host stars could have formed {\em in situ}. They encountered difficulties forming the planets within the required time scales of estimated disk lifetimes unless certain nebular parameters were set to unlikely values, but cautioned that there are still major uncertainties associated with the nebula model. Thus it seems more likely that gas planets form at large distances ($\gtrsim$ 5~AU) and migrate towards their observed positions, a process possibly caused by disk-planet interactions \citep[see, e.g.,][ and references therein]{Papa:99,Tril:02}.
On the other hand, \citet{Ikom:01} do not exclude the possibility of in situ formation of extrasolar giant planets.
\citet{Boss:02} calculated disk instability models with a factor of 10 lower and higher than solar opacities and showed that all models formed clumps at distances between $\approx$ 5 and 10~AU at similar time scales. Similarly, \citet{Ikom:00} found that within the core instability model the growth time of the gas envelope masses depend only moderately on grain opacity.

Thus, at the moment it is not clear how the intrinsic metallicity of the protoplanetary disk influences giant planet formation. On the other hand, it has been proposed that the stellar abundances might be altered as a result of planet formation. This possibility has been explored by, e.g., \citet{Sand:02}, and \citet{Murr:02}, who calculated the metallicities of stellar models with different amounts of iron added to their envelopes. This material is assumed to come from planets that have fallen into the central star after their formation. Since the amount by which the heavy element abundance can be changed by such a process depends mainly on the mass of the mixed layer of the atmosphere, i.e. the convection zone, the effect is expected to be largest for high temperature, high mass stars, which have least massive convection zones. Our sample of stars shows the opposite behavior: the abundances of the coolest CGP dwarfs are most enhanced compared to the no-CGP dwarfs (Fig.~\ref{results_bin}). However, the fact that the metallicity differences do vary with temperature in a nontrivial way could point to an influence of the planet formation process.

One aspect of planetary system formation which has not been discussed up to now in connection with metallicity is the possible accretion of gas from the planetary disk onto the star. Accretion disks have been observed around T~Tauri stars \citep[e.g.][]{Hart:98} and have been modeled by, e.g., \citet{Armi:01}, with magnetohydrodynamic turbulence as the source for angular momentum transport.
The presence of accretion disks could alter the stellar surface abundance if we consider the following assumptions:
(1) accretion is still taking place at a significant rate when solid bodies large enough to be decoupled from the gas have formed, (2) the convection zone is already small enough so that abundances within it are changed by adding material, (3) the accreted gas is depleted in heavy elements, because they are locked up in the solid bodies. We should then expect the heavy element abundance to {\em decrease} with respect to the initial stellar/disk abundance.

We made rough estimates of the metallicity change for various values of the initial abundance, the mass of the convection zone, the amount of heavy elements contained in solid bodies (which determines the abundance of the accreted gas), and the accreted gas mass (which should depend mainly on the mass of the disk and the amount of mass lost from the disk).
It became clear that a measurable effect is likely to occur only in the most massive young stars, for which the convection zone mass might be as low as $\approx$0.03\Msun\ \citep[c.f.][ Fig. 1]{Murr:01}. But these are the stars for which the average metallicity is problematic to explain within the ``pollution by planets'' scenario, and one solution could be that the accretion of depleted disk gas is more important than planet accretion.
The decrease in abundance is smallest (negligible) for high initial abundances ([Fe/H]$_{\rm i}\gtrsim$+0.3), and of course for low accreted mass (e.g. 0.005~\Msun) and if only a small amount of high-Z material is assumed to be in solid bodies (20-40~\Me, corresponding to 1-2~\Me\ of iron, for solar abundances).
Assuming an accreted mass of 0.01\Msun\ \citep[e.g. accretion at a rate of $10^{-8}$~\Msun~yr$^{-1}$ for $10^6$~yr, cf.][]{Hart:98} of gas with heavy elements (Z) reduced by 80~\Me\ (5~\Me\ of iron), we estimate the decrease in abundance to be on the order of 0.1~dex for $-$0.2$\lesssim$[Fe/H]$_{\rm i}\lesssim$0.0~dex, and 0.05~dex for 0.0$\lesssim$[Fe/H]$_{\rm i}\lesssim$+0.3~dex.
These estimated changes are small compared to the observed scatter in abundances, but they involve several not very well constrained parameters and detailed models would be necessary to assess their importance. As abundance determinations are becoming more and more accurate these effects will have to be considered.

When we add planet migration to this scenario and as a consequence accretion of a part of the solid bodies onto the star, we expect to obtain abundances which are enhanced compared to [Fe/H]$_{\rm i}$. The enhancement again is estimated to depend strongly on [Fe/H]$_{\rm i}$. For example, adding 30\% of the mass contained in solid bodies to the stellar atmosphere in the case described above, now assuming that the convection zone has shrunk to 0.006~\Msun, should result in an enhancement of 0.25~dex for [Fe/H]$_{\rm i}\approx-$0.2~dex and of 0.1~dex for [Fe/H]$_{\rm i}\approx+$0.3~dex. In summary, depending on the initial metallicity and the efficiency of gas and solids accretion, the more massive dwarfs with planets should show a variety of abundances. But to study the metallicity distribution of massive dwarfs with planets in detail would require more measurements as well as more than the above rule of thumb calculations.

A final point to be discussed in this context is the paucity of no-CGP dwarfs with high metallicity. Each of these stars could have planets with periods beyond the current detection limits. In that case it is also conceivable that they experienced accretion as described above and therefore they should include high-metallicity stars. There are two stars (of very different temperatures) in our sample which fit this description. Two other no-CGP dwarfs had also been classified as ``VSL'', but their metallicities turned out to be nearly solar. The low metallicities obtained for some of the no-CGP dwarfs could be a result of accretion of depleted disk gas only (no accretion of planetary bodies).

In this discussion, we considered only the influence of an addition of material with solar scaled heavy element abundance changes on [Fe/H], while different elements (in particular refractory vs. volatile elements) are expected to be affected differently. Results for twelve stars studied by \citet{Sada:02} indicate that this is not the case.
On the other hand, \citet{Smit:01} found larger than average slopes of element abundance versus condensation temperature for five out of 30 planet host stars \citep[most of them from][]{Gonz:01}. Furthermore, \citet{Grat:01} found differences in element abundances for two non-interacting main sequence binaries which might suggest infall of rocky material.
Abundance determination for elements other than iron are planned for our program stars to further clarify the situation.

\section{Summary}

We have obtained high resolution spectra for three samples of dwarf stars related to extrasolar planetary systems: stars which are known to host close-in giant planets (``CGP dwarfs''), stars for which radial velocity data exclude the presence of a close-in giant planetary companion (``no-CGP dwarfs'') and a random sample of dwarfs with a spectral type and magnitude distribution similar to that of the former two groups (``control dwarfs'').
In addition, we observed a sample of ``Very Strong Lined'' or VSL dwarfs, which brings the total number of stars presented in this paper to 114 (140 when adding the stars in Appendix A).
All stars have been observed with the same telescope/instrument and have been analyzed using the same model atmospheres, atomic data and equivalent width modeling program.
We derived the \Teff, \logg, \vmicro\ and [Fe/H] values for each star which give a best fit to the observed equivalent widths, using a large number of Fe lines.
Line abundances have been derived differentially to the Sun, using a solar spectrum obtained with Callisto as the reflector with the same instrumentation.

The results can be summarized as follows:
\begin{itemize}
\item Taking the difference between stellar and solar abundance for each spectral line decreases the scatter of the line abundances around the mean significantly and we can use several hundred lines per star for the analysis.
\item A comparison of the parameters derived in this work (spec) with that obtained from Geneva (G) or Str\"omgren (S) photometric calibrations reveals systematic differences: either a largely constant offset towards lower values -- 0.08~dex for [Fe/H](G) and 105~K for \Teff(S) -- or a difference which depends on the value of the parameter, with slope $\approx$0.8 in the \Teff(spec) vs. \Teff(G) and the [Fe/H](spec) vs. [Fe/H](S) diagrams.
\item A comparison of the parameters derived in this work with that obtained by other authors from high resolution spectroscopy shows that the temperatures agree well except for the coolest and/or most metal rich stars (the mean difference is 85~$\pm$120~K). The iron abundances agree very well (the mean difference is 0.03~$\pm$0.07~dex), except for five stars.
\item The mean iron abundance of the control dwarf sample is solar ($\sigma$=0.4~dex).
The iron abundances of CGP dwarfs are on average by 0.22~dex greater than that of no-CGP dwarfs.
\item The iron abundance distributions of both the CGP and no-CGP dwarfs are different than that of the control dwarfs, while the combined iron abundances have a distribution which is very similar to that of the control dwarfs.
\item All four samples (CGP, no-CGP, combined, control) have very different temperature distributions. The most significant difference lies in the number of dwarfs with low temperatures -- 16 no-CGP and five CGP dwarfs with \Teff $<$ 5600~K.
\item Combining the temperature and metallicity information, we show that metal enrichment occurs only for CGP dwarfs with temperatures just below solar and $\approx$ 300~K higher than solar, whereas there is no difference at \Teff\ around 6000~K.
\item About 3/4 of the stars classified as VSL have [Fe/H] $\ge$ +0.1.
\end{itemize}

The metallicity differences between CGP and no-CGP dwarfs could either be caused by a difference in the intrinsic abundance distributions or by accretion of material during the planet formation process. The influence of the intrinsic metallicity of the protoplanetary disk on giant planet formation has not been well established up to now. On the other hand, it has been shown that the effect of ``pollution'' would be largest for high temperature stars and decrease with temperature. Since we observe the opposite behavior, we conclude that the 
intrinsic metallicity distribution seems to be more important than planet accretion. For some of the most massive stars, the effect of ``pollution'' could be mitigated by preceding accretion of heavy element depleted disk gas.

Planned future work includes establishing improved calibrations for atmospheric parameters using line-depth ratios or photometry, as well as the determination of abundances for elements other than iron.

\acknowledgments

This research is being funded by NSF grant AST-0086249.
We acknowledge the financial support of CWRU which has made it possible to obtain the observations used in this work.
Use was made of the Simbad database, operated at CDS, Strasbourg, France, and of NASA's Astrophysics Data System Bibliographic Services.
We thank the referee for helpful comments.

\appendix

\section{Additional Planet Hosts}

Abundance analysis results for additional planet hosts observed by us are given in Table~\ref{param_tab_a}. The metallicity distribution of the stars hosting planets with minimum masses below the brown dwarf mass limit seems similar to that of the CGP dwarfs. Note that the seven brown dwarf hosts span a large range (about an order of magnitude) in [Fe/H].
For BD\,-10\,3166 we derive an abundance which is by 0.18~dex higher than that given by \citet{Gonz:01}. The difference reduces to +0.13~dex if we use the the parameters of \citet{Gonz:01}, but we obtain significant trends of Fe~I abundance with equivalent width and excitation energy, and the difference between Fe~I and Fe~II is $-$0.24~dex. This result does not change if in addition we use only the lines from their line list. The remaining difference might well be due to differences in the measured equivalent widths -- our equivalent widths are on average 14\% higher than that of \citet{Gonz:01}. This indicates that we have estimated the continuum flux to lie at a higher level for this faint, metal rich star.

Table~\ref{param_tab_a} includes the second component (B) of the wide binary 16\,Cyg (HD\,186427), which hosts a planet with too low a mass (for its semi-major axis) and too high an eccentricity to be selected according to our criteria, whereas radial velocity data exclude planets for 16\,Cyg\,A (HD\,186408) at the chosen level, which is therefore included in the no-CGP list (Table~\ref{param_tab_2}). The two stars have very similar properties, and within the errors their iron abundances are equal. However, formally we find exactly the same difference in iron abundance between A and B as, e.g. \citet{Laws:01}, i.e. +0.03~dex. We do not want to draw any conclusions from this finding here, because in accordance with our classification, 16\,Cyg\,A could have an as yet undetected planet similar to that of 16\,Cyg\,B. 


\bibliographystyle{apj} 
\bibliography{apj-jour,nearbystars}


\clearpage

\clearpage

\clearpage

\clearpage

\clearpage

\clearpage

\clearpage

\clearpage

\clearpage

\clearpage

\clearpage

\clearpage

\clearpage

\clearpage

\clearpage

\clearpage

\clearpage

\clearpage

\begin{deluxetable}{rccrrrp{3mm}ccrrr}
\tabletypesize{\scriptsize}
\tablecaption{Observing log for the solar spectrum (Callisto) and the CGP dwarf spectra. The columns list HD number, central wavelength of the central echelle order, date of observation, integration time in seconds, average signal-to-noise per pixel for the central order, and observer (L\dots R.E. Luck, H\dots U. Heiter). \label{obs_log_dwp}}
\tablewidth{0pt}
\tablehead{
 & \multicolumn{5}{c}{blue spectrum} & & \multicolumn{5}{c}{red spectrum} \\ 
\colhead{HD} & \colhead{$\lambda_{\rm c}$ [\AA]}  & \colhead{Date} & \colhead{$t_{\rm i}$ [s]} & \colhead{S/N}  & \colhead{O} &  &  \colhead{$\lambda_{\rm c}$ [\AA]}  & \colhead{Date} & \colhead{$t_{\rm i}$ [s]} & \colhead{S/N}  & \colhead{O}
}
\startdata
Callisto   &  5276 &   2001-10-26 &   750 &   226 & L & &  6263 &   2001-10-28 &   900 &   201 & L \\
8574       &  5205 &   2001-08-23 &  1800 &   228 & L & &  6212 &   2001-08-27 &   900 &   219 & L \\
9826       &  5241 &   1999-10-19 &    90 &   308 & L & &  6213 &   1999-10-23 &    60 &   338 & L \\
 &  5241 &   1999-10-19 &    90 &   289 & L & & & & & & \\
19994      &  5205 &   2001-08-23 &   300 &   102 & L & &  6162 &   2000-10-20 &   300 &   156 & L \\
 &  5205 &   2001-08-25 &   180 &   185 & L & &  6211 &   2001-02-11 &   600 &   319 & L \\
23596      &  5241 &   2002-11-14 &  2700 &   204 & H  & &  6263 &   2002-11-18 &  2400 &   279 & H  \\
28185      &  5275 &   2001-10-25 &  1800 &   194 & L & &  6262 &   2001-10-27 &  1200 &   214 & L \\
33636      &  5276 &   2002-01-23 &  1800 &   224 & H  & &  6262 &   2002-01-26 &   900 &   193 & H  \\
38529      &  5202 &   1997-01-31 &   900 &   273 & L & &  6363 &   1997-02-02 &   600 &   360 & L \\
40979      &  5240 &   2002-11-14 &  1800 &   212 & H  & &  6262 &   2002-11-18 &  1500 &   262 & H  \\
50554      &  5276 &   2001-10-23 &   900 &   111 & L & &  6263 &   2001-10-27 &   900 &   275 & L \\
52265      &  5275 &   2001-10-24 &  1200 &   265 & L & &  6262 &   2001-10-27 &   900 &   286 & L \\
68988      &  5277 &   2002-01-24 &  2700 &    99 & H  & &  6264 &   2002-01-26 &  2400 &   163 & H  \\
72659      &  5241 &   2002-11-14 &  2700 &   183 & H  & &  6264 &   2002-11-19 &  2100 &   239 & H  \\
75732      &  5240 &   1999-10-19 &   600 &   220 & L & &  6363 &   1997-02-03 &   420 &   200 & L \\
92788      &  5204 &   2001-05-08 &  2700 &   195 & L & &  6065 &   2001-05-14 &  1200 &   210 & L \\
114762     &  5134 &   1999-05-28 &  1800 &   181 & L & &  6017 &   1999-05-26 &  1800 &   340 & L \\
& & & & & &  &  6262 &   2000-01-25 &  1200 &   261 & L \\
117176     &  5135 &   1999-05-28 &   720 &   288 & L & &  6017 &   1999-05-25 &   600 &   467 & L \\
& & & & & &  &  6017 &   1999-05-25 &   420 &   382 & L \\
& & & & & &  &  6263 &   2000-01-25 &   120 &   261 & L \\
120136     &  5135 &   1999-05-28 &   600 &   299 & L & &  6018 &   1999-05-25 &   300 &   309 & L \\
& & & & & &  &  6263 &   2000-01-25 &   100 &   302 & L \\
128311     &  5241 &   2003-02-18 &  2700 &   197 & H  & &  6213 &   2003-02-22 &  2100 &   247 & H  \\
130322     &  5241 &   2000-01-30 &  1800 &   174 & L & &  6263 &   2000-01-26 &  1800 &   195 & L \\
134987     &  5205 &   2000-04-30 &  1000 &   226 & L & &  6366 &   2000-04-25 &   450 &   281 & L \\
136118     &  5240 &   2002-05-21 &  2100 &   203 & L & &  6212 &   2002-05-24 &  1800 &   225 & L \\
141937     &  5205 &   2001-05-08 &  2700 &   180 & L & &  6264 &   2002-01-26 &  2400 &   160 & H  \\
143761     &  5135 &   1999-05-28 &   600 &   299 & L & &  6017 &   1999-05-26 &   400 &   400 & L \\
 &  5240 &   2001-05-10 &   600 &   301 & L & & & & & & \\
145675     &  5385 &   1998-05-15 &  1800 &   258 & L & &  6312 &   1997-04-24 &   750 &   240 & L \\
169830     &  5135 &   2000-08-21 &   600 &   186 & L & &  6113 &   2000-08-18 &   450 &   253 & L \\
177830     &  5136 &   2000-08-21 &  1800 &   238 & L & &  6115 &   2000-08-18 &   600 &   218 & L\\
178911     &  5205 &   2001-05-08 &  2700 &   138 & L & &  6213 &   2001-08-27 &  2400 &   166 & L \\
179949     &  5205 &   2001-08-24 &   600 &   192 & L & &  6212 &   2001-08-28 &   600 &   258 & L \\
187123     &  5136 &   1999-05-30 &  2700 &   163 & L & &  6212 &   1999-10-25 &  1800 &   286 & L \\
190228     &  5241 &   2000-10-22 &  1200 &   177 & L & &  6213 &   2001-08-27 &  1628 &   181 & L \\
 &  5205 &   2001-08-23 &  1200 &   153 & L & & & & & & \\
195019     &  5137 &   1999-05-30 &  1800 &   235 & L & &  6214 &   1999-10-25 &   900 &   331 & L \\
209458     &  5135 &   2000-08-21 &  1800 &   189 & L & &  6114 &   2000-08-18 &  1200 &   231 & L \\
217014     &  5241 &   1999-10-19 &   300 &   267 & L & &  6161 &   1997-10-18 &   300 &   302 & L \\
217107     &  5240 &   1999-10-19 &   600 &   257 & L & &  6212 &   1999-10-24 &   420 &   330 & L \\
\enddata
\end{deluxetable}

\clearpage

\begin{deluxetable}{rccrrrp{3mm}ccrrr}
\tabletypesize{\scriptsize}
\tablecaption{Observing log for the no-CGP dwarf spectra. For column explanation see Table~\ref{obs_log_dwp}. \label{obs_log_dwop}}
\tablewidth{0pt}
\tablehead{
 & \multicolumn{5}{c}{blue spectrum} & & \multicolumn{5}{c}{red spectrum} \\ 
\colhead{HD} & \colhead{$\lambda_{\rm c}$ [\AA]}  & \colhead{Date} & \colhead{$t_{\rm i}$ [s]} & \colhead{S/N}  & \colhead{O} &  &  \colhead{$\lambda_{\rm c}$ [\AA]}  & \colhead{Date} & \colhead{$t_{\rm i}$ [s]} & \colhead{S/N}  & \colhead{O}
}
\startdata
166        &  5135 &   2000-08-22 &   600 &   178 &    L &  &  6212 &   1999-10-25 &   600 &   317 &    L\\
4614       &  5135 &   2000-08-22 &   120 &   244 &    L &  &  6212 &   1999-10-25 &    60 &   342 &    L\\
4628       &  5136 &   2000-08-22 &   600 &   210 &    L &  &  6163 &   1999-10-25 &   900 &   392 &    L\\
 & & & & &  &  &  6114 &   2000-08-20 &   360 &   255 &    L\\
10476      &  5206 &   2001-08-23 &   300 &   114 &    L &  &  6114 &   2000-08-20 &   240 &   267 &    L\\
10700      &  5241 &   1999-10-20 &    75 &   318 &    L &  &  6212 &   1999-10-24 &    35 &   281 &    L\\
 & & & & &  &  &  6212 &   1999-10-24 &    40 &   339 &    L\\
12235      &  5241 &   1999-10-19 &   600 &   259 &    L &  &  6364 &   1997-02-03 &   600 &   312 &    L\\
16160      &  5205 &   2001-08-25 &   450 &   177 &    L &  &  6113 &   2000-08-18 &   600 &   386 &    L\\
 & & & & &  &  &  6113 &   2000-08-18 &   300 &   288 &    L\\
16895      &  5276 &   2001-10-23 &   180 &   231 &    L &  &  6113 &   2000-08-19 &   120 &   255 &    L\\
22484      &  5240 &   1999-10-21 &    90 &   260 &    L &  &  6212 &   1999-10-24 &    75 &   295 &    L\\
25680      &  5240 &   2002-11-15 &  1200 &   231 &    H &  &  6212 &   2003-02-23 &   600 &   255 &    H\\
26965      &  5241 &   1999-10-19 &   120 &   268 &    L &  &  6213 &   1999-10-23 &    60 &   290 &    L\\
32147      &  5202 &   1997-01-31 &   600 &   164 &    L &  &  6363 &   1997-02-02 &   600 &   306 &    L\\
48682      &  5277 &   2001-10-25 &   600 &   353 &    L &  &  6163 &   2000-10-20 &   450 &   149 &    L\\
 & & & & &  &  &  6264 &   2001-10-27 &   500 &   310 &    L\\
50281      &  5277 &   2001-10-26 &   900 &   199 &    L &  &  6263 &   2002-01-25 &  1800 &   268 &    H\\
52711      &  5240 &   2002-11-15 &   900 &   254 &    H &  &  6263 &   2002-11-18 &   900 &   242 &    H\\
61421      &  5240 &   2000-01-29 &     6 &   341 &    L &  &  6263 &   2000-01-26 &     8 &   404 &    L\\
69830      &  5240 &   2003-02-18 &   720 &   257 &    H &  &  6262 &   2002-01-25 &  1200 &   295 &    H\\
69897      &  5240 &   2002-11-15 &   450 &   269 &    H &  &  6263 &   2002-11-18 &   420 &   222 &    H\\
76151      &  5240 &   2000-01-29 &  1200 &   262 &    L &  &  6363 &   1997-02-03 &   600 &   380 &    L\\
78366      &  5239 &   2002-05-23 &  1200 &   140 &    L &  &  6211 &   2002-05-24 &   600 &   289 &    L\\
82328      &  5135 &   1999-05-30 &    60 &   287 &    L &  &  6017 &   1999-05-26 &    50 &   352 &    L\\
84737      &  5240 &   2002-05-21 &   360 &   203 &    L &  &  6212 &   2002-05-24 &   360 &   302 &    L\\
90839      &  5135 &   1999-05-30 &   480 &   346 &    L &  &  6017 &   1999-05-26 &   250 &   356 &    L\\
126053     &  5240 &   2002-05-21 &  1200 &   176 &    L &  &  6212 &   2002-05-24 &  1200 &   256 &    L\\
149661     &  5275 &   1997-04-28 &   900 &   298 &    L &  &  6312 &   1997-04-24 &   750 &   327 &    L\\
155886     &  5135 &   1999-05-28 &   600 &   257 &    L &  &  6018 &   1999-05-26 &   400 &   421 &    L\\
157214     &  5206 &   2000-04-29 &   900 &   284 &    L &  &  6368 &   2000-04-25 &   200 &   289 &    L\\
166620     &  5205 &   2001-08-25 &   500 &   202 &    L &  &  6212 &   2001-08-27 &   500 &   253 &    L\\
 & & & & &  &  &  6162 &   2000-10-19 &   180 &   133 &    L\\
170657     &  5205 &   2001-08-24 &   600 &   145 &    L &  &  6212 &   2001-08-27 &   600 &   126 &    L\\
185144     &  5240 &   1999-10-19 &   120 &   237 &    L &  &  6212 &   1999-10-23 &   120 &   307 &    L\\
186408     &  5136 &   1999-05-28 &   720 &   220 &    L &  &  6019 &   1999-05-26 &   720 &   331 &    L\\
 & & & & &  &  &  6114 &   2000-08-19 &   420 &   243 &    L\\
201091     &  5241 &   1999-10-19 &   250 &   272 &    L &  &  6162 &   1997-08-17 &   180 &   426 &    L\\
215648     &  5240 &   2002-08-27 &   120 &   187 &    H &  &  6212 &   2002-08-31 &    90 &   230 &    H\\
219134     &  5241 &   1999-10-19 &   300 &   277 &    L &  &  6213 &   1999-10-24 &   250 &   341 &    L\\
222368     &  5135 &   2000-08-21 &   180 &   236 &    L &  &  6113 &   2000-08-19 &   120 &   275 &    L\\
\enddata
\end{deluxetable}

\clearpage

\begin{deluxetable}{rccrrrp{3mm}ccrrr}
\tabletypesize{\scriptsize}
\tablecaption{Observing log for the control dwarf spectra. For column explanation see Table~\ref{obs_log_dwp}. \label{obs_log_c}}
\tablewidth{0pt}
\tablehead{
 & \multicolumn{5}{c}{blue spectrum} & & \multicolumn{5}{c}{red spectrum} \\ 
\colhead{HD} & \colhead{$\lambda_{\rm c}$ [\AA]}  & \colhead{Date} & \colhead{$t_{\rm i}$ [s]} & \colhead{S/N}  & \colhead{O} &  &  \colhead{$\lambda_{\rm c}$ [\AA]}  & \colhead{Date} & \colhead{$t_{\rm i}$ [s]} & \colhead{S/N}  & \colhead{O}
}
\startdata
5015       &  5135 &   2000-08-22 &   300 &   255 &    L &  &  6212 &   1999-10-25 &   300 &   380 &    L\\
10780      &  5135 &   2000-08-22 &   600 &   180 &    L &  &  6212 &   1999-10-25 &   450 &   319 &    L\\
25998      &  5240 &   1999-10-20 &   200 &   228 &    L &  &  6212 &   1999-10-23 &   220 &   294 &    L\\
46588      &  5276 &   2001-10-25 &   600 &   307 &    L &  &  6162 &   2000-10-20 &   600 &   215 &    L\\
 & & & & &  &  &  6263 &   2001-10-27 &   500 &   306 &    L\\
72945      &  5203 &   2000-04-30 &   600 &   273 &    L &  &  6263 &   2000-01-26 &   450 &   242 &    L\\
72946      &  5204 &   2000-04-30 &  1200 &   185 &    L &  &  6262 &   2000-01-26 &  1200 &   220 &    L\\
81858      &  5240 &   2000-01-31 &   300 &   246 &    L &  &  6263 &   2000-01-26 &   300 &   238 &    L\\
82885      &  5240 &   2000-01-31 &   300 &   262 &    L &  &  6262 &   2000-01-26 &   300 &   327 &    L\\
 &  5204 &   2000-04-30 &   420 &   274 &    L &  &  6365 &   2000-04-25 &   300 &   318 &    L\\
85380      &  5240 &   2000-01-31 &   600 &   227 &    L &  &  6262 &   2000-01-28 &   600 &   266 &    L\\
88595      &  5240 &   2000-01-30 &   900 &   225 &    L &  &  6262 &   2000-01-28 &   600 &   221 &    L\\
101177     &  5241 &   2000-01-29 &  1800 &   219 &    L &  &  6364 &   1997-02-03 &   420 &   267 &    L\\
101501     &  5204 &   2000-04-30 &   420 &   335 &    L &  &  6364 &   1997-02-03 &   240 &   354 &    L\\
 &  5240 &   2000-01-29 &   900 &   283 &    L &  &  6366 &   2000-04-25 &   300 &   435 &    L\\
 & & & & &  &  &  6366 &   2000-04-25 &   180 &   335 &    L\\
101563     &  5240 &   2001-05-10 &   600 &   135 &    L &  &  6365 &   1997-02-03 &   360 &   219 &    L\\
103095     &  5242 &   2000-01-29 &  1800 &   193 &    L &  &  6366 &   1997-02-03 &   420 &   302 &    L\\
109358     &  5240 &   2000-01-29 &   420 &   289 &    L &  &  6364 &   1997-02-03 &    60 &   290 &    L\\
 & & & & &  &  &  6263 &   2000-01-26 &   120 &   327 &    L\\
110897     &  5239 &   2000-01-30 &   600 &   291 &    L &  &  6363 &   1997-02-03 &   420 &   329 &    L\\
111395     &  5241 &   2000-01-30 &   900 &   298 &    L &  &  6365 &   1997-02-03 &   420 &   307 &    L\\
126141     &  5205 &   2000-04-30 &   900 &   311 &    L &  &  6366 &   2000-04-25 &   400 &   275 &    L\\
143333     &  5205 &   2000-04-29 &   900 &   237 &    L &  &  6367 &   2000-04-25 &   300 &   374 &    L\\
176051     &  5206 &   2000-04-30 &   600 &   306 &    L &  &  6114 &   2000-08-19 &   150 &   260 &    L\\
179957     &  5136 &   2000-08-21 &  1200 &   197 &    L &  &  6114 &   2000-08-20 &   900 &   238 &    L\\
179958     &  5136 &   2000-08-21 &  1200 &   199 &    L &  &  6114 &   2000-08-20 &   900 &   257 &    L\\
184151     &  5135 &   2000-08-22 &   600 &   176 &    L &  &  6113 &   2000-08-20 &  1200 &   261 &    L\\
187691     &  5135 &   2000-08-22 &   240 &   253 &    L &  &  6113 &   2000-08-20 &   360 &   327 &    L\\
 & & & & &  &  &  6113 &   2000-08-20 &   240 &   279 &    L\\
188376     &  5135 &   2000-08-22 &   300 &   205 &    L &  &  6113 &   2000-08-20 &   180 &   218 &    L\\
193555     &  5135 &   2000-08-22 &  1200 &   253 &    L &  &  6114 &   2000-08-19 &   600 &   205 &    L\\
193664     &  5135 &   2000-08-22 &   720 &   254 &    L &  &  6113 &   2000-08-19 &   450 &   213 &    L\\
197963     &  5135 &   2000-08-22 &   420 &   260 &    L &  &  6113 &   2000-08-18 &   180 &   229 &    L\\
216172     &  5135 &   2000-08-22 &   900 &   182 &    L &  &  6113 &   2000-08-19 &   600 &   132 &    L\\
216172     &  5135 &   2000-08-22 &  1200 &   180 &    L &  &  6113 &   2000-08-19 &  1200 &   180 &    L\\
221445     &  5135 &   2000-08-21 &  1200 &   160 &    L &  &  6114 &   2000-08-19 &   900 &   186 &    L\\
224930     &  5205 &   2001-08-24 &   600 &   254 &    L &  &  6114 &   2000-08-20 &   360 &   252 &    L\\
\enddata
\end{deluxetable}

\clearpage

\begin{deluxetable}{rccrrrp{3mm}ccrrr}
\tabletypesize{\scriptsize}
\tablecaption{Observing log for the VSL dwarf spectra. For column explanation see Table~\ref{obs_log_dwp}. \label{obs_log_vsl}}
\tablewidth{0pt}
\tablehead{
 & \multicolumn{5}{c}{blue spectrum} & & \multicolumn{5}{c}{red spectrum} \\ 
\colhead{HD} & \colhead{$\lambda_{\rm c}$ [\AA]}  & \colhead{Date} & \colhead{$t_{\rm i}$ [s]} & \colhead{S/N}  & \colhead{O} &  &  \colhead{$\lambda_{\rm c}$ [\AA]}  & \colhead{Date} & \colhead{$t_{\rm i}$ [s]} & \colhead{S/N}  & \colhead{O}
}
\startdata
1461       &  5240 &   1999-10-19 &   750 &   227 &    L &  &  6212 &   1999-10-23 &   300 &   254 &    L\\
9562       &  5240 &   1999-10-19 &   500 &   251 &    L &  &  6212 &   1999-10-23 &   250 &   292 &    L\\
20630      &  5202 &   1997-01-30 &  1200 &   303 &    L &  &  6363 &   1997-02-02 &   240 &   346 &    L\\
73752      &  5239 &   2000-01-29 &   600 &   248 &    L &  &  6363 &   1997-02-03 &   300 &   326 &    L\\
99491      &  5203 &   1997-01-29 &   600 &   188 &    L &  &  6364 &   1997-02-01 &   750 &   246 &    L\\
 & & & & &  &  &  6017 &   1999-05-25 &   900 &   230 &    L\\
99492      &  5203 &   1997-01-29 &  1500 &   124 &    L &  &  6364 &   1997-02-01 &  1800 &   236 &    L\\
 & & & & &  &  &  6017 &   1999-05-25 &  1800 &   242 &    L\\
104304     &  5385 &   1998-05-16 &  1800 &   242 &    L &  &  6017 &   1999-05-25 &   600 &   322 &    L\\
114710     &  5385 &   1998-05-16 &   900 &   410 &    L &  &  6017 &   1999-05-25 &   150 &   347 &    L\\
115617     &  5385 &   1998-05-16 &  1200 &   176 &    L &  &  6018 &   1999-05-25 &   200 &   271 &    L\\
125184     &  5385 &   1998-05-16 &  2204 &   136 &    L &  &  6018 &   1999-05-26 &   900 &   336 &    L\\
135101     &  5275 &   1997-04-28 &  2400 &   303 &    L &  &  6365 &   1997-02-02 &   900 &   330 &    L\\
 &  5275 &   1997-04-29 &   900 &   203 &    L &  &  6313 &   1997-04-23 &   900 &   362 &    L\\
 &  5386 &   1998-05-16 &  2700 &   191 &    L &  &  6018 &   1999-05-26 &  1000 &   348 &    L\\
135101     &  5275 &   1997-04-28 &  3600 &   251 &    L &  &  6365 &   1997-02-02 &  1800 &   305 &    L\\
 &  5275 &   1997-04-29 &  2400 &   218 &    L &  &  6313 &   1997-04-23 &  1800 &   353 &    L\\
 &  5386 &   1998-05-16 &  2700 &   152 &    L &  &  6018 &   1999-05-26 &  1800 &   325 &    L\\
182572     &  5387 &   1998-05-15 &  1200 &   266 &    L &  &  6162 &   1997-10-18 &   300 &   296 &    L\\
\enddata
\end{deluxetable}

\clearpage

\begin{deluxetable}{rrrrrrrr}
\tabletypesize{\footnotesize}
\tablecaption{Best-fit parameters and iron abundances for the CGP dwarfs. The units are K for \Teff, cgs-units for \logg, and \kms\ for \vmicro. \label{param_tab_1}}
\tablewidth{0pt}
\tablehead{
\colhead{HD} & \colhead{\Teff} & \colhead{\logg} & \colhead{\vmicro} & \colhead{[Fe/H]} & \colhead{$\sigma_{\rm [Fe/H]}$} & \colhead{$N_{\rm I}$} & \colhead{$N_{\rm II}$}
}
\startdata
8574 & 6100 & 4.30 & 1.45 & $-$0.04 & 0.05 & 349 & 28\\
9826 & 6200 & 4.40 & 1.60 & 0.13 & 0.07 & 353 & 24\\
19994 & 6150 & 4.30 & 1.50 & 0.23 & 0.07 & 378 & 24\\
23596\tablenotemark{1} & 6150 & 4.45 & 1.45 & 0.28 & 0.05 & 425 & 25\\
28185 & 5700 & 4.20 & 0.80 & 0.24 & 0.06 & 454 & 22\\
33636\tablenotemark{1} & 6050 & 4.55 & 1.25 & $-$0.11 & 0.05 & 360 & 25\\
38529\tablenotemark{v} & 5700 & 4.00 & 1.45 & 0.40 & 0.08 & 407 & 21\\
40979 & 6200 & 4.60 & 1.25 & 0.19 & 0.06 & 397 & 23\\
50554 & 6050 & 4.50 & 1.20 & $-$0.04 & 0.04 & 376 & 24\\
52265 & 6100 & 4.40 & 1.30 & 0.15 & 0.05 & 410 & 27\\
68988 & 6000 & 4.45 & 1.35 & 0.36 & 0.06 & 420 & 25\\
72659 & 5950 & 4.35 & 1.30 & $-$0.07 & 0.04 & 376 & 21\\
75732\tablenotemark{v} & 5500 & 4.40 & 1.00 & 0.46 & 0.09 & 441 & 21\\
92788\tablenotemark{2} & 5700 & 4.20 & 0.50 & 0.30 & 0.06 & 385 & 24\\
114762\tablenotemark{1} & 6000 & 4.40 & 1.50 & $-$0.78 & 0.06 & 245 & 12\\
117176\tablenotemark{1} & 5600 & 4.10 & 1.00 & $-$0.05 & 0.03 & 392 & 23\\
120136 & 6600 & 4.70 & 1.90 & 0.37 & 0.12 & 317 & 19\\
128311 & 5200 & 4.50 & 1.15 & 0.13 & 0.11 & 399 & 15\\
130322 & 5400 & 4.40 & 0.00 & 0.10 & 0.06 & 462 & 22\\
134987 & 5900 & 4.40 & 1.30 & 0.34 & 0.06 & 412 & 24\\
136118\tablenotemark{1} & 6250 & 4.45 & 1.60 & $-$0.03 & 0.07 & 335 & 23\\
141937\tablenotemark{1} & 5900 & 4.45 & 0.90 & 0.12 & 0.04 & 412 & 22\\
143761\tablenotemark{3} & 5900 & 4.40 & 1.25 & $-$0.25 & 0.04 & 314 & 22\\
145675\tablenotemark{v} & 5600 & 4.40 & 1.20 & 0.54 & 0.10 & 393 & 22\\
169830\tablenotemark{2} & 6300 & 4.40 & 1.60 & 0.09 & 0.06 & 341 & 28\\
177830\tablenotemark{2} & 5000 & 3.60 & 0.90 & 0.57 & 0.11 & 275 & 7 \\
178911 & 5600 & 4.20 & 0.30 & 0.31 & 0.07 & 408 & 26\\
179949 & 6200 & 4.50 & 1.20 & 0.20 & 0.06 & 369 & 24\\
187123 & 5800 & 4.35 & 0.90 & 0.08 & 0.04 & 384 & 23\\
190228\tablenotemark{2} & 5360 & 3.90 & 0.90 & $-$0.17 & 0.04 & 411 & 23\\
195019\tablenotemark{3} & 5830 & 4.30 & 1.05 & 0.05 & 0.04 & 384 & 25\\
209458 & 6100 & 4.50 & 1.30 & $-$0.02 & 0.05 & 320 & 22\\
217014\tablenotemark{v} & 5750 & 4.25 & 0.65 & 0.23 & 0.05 & 417 & 23\\
217107 & 5750 & 4.35 & 1.15 & 0.39 & 0.06 & 412 & 24\\
\enddata
\tablenotetext{1}{Companion with $M\sin i \ge 6.5~M_{\rm J}$}
\tablenotetext{2,3}{Companion might be brown dwarf candidate (2) or M dwarf (3), according to \citet{Han:01}}
\tablenotetext{v}{VSL dwarf}
\end{deluxetable}

\clearpage

\begin{deluxetable}{rrrrrrrr}
\tabletypesize{\footnotesize}
\tablecaption{Best-fit parameters and iron abundances for the no-CGP dwarfs. The no-CGP dwarfs have been selected from the list of \citet{Cumm:99}, unless otherwise noted. Units are as in Table~\ref{param_tab_1}. \label{param_tab_2}}
\tablewidth{0pt}
\tablehead{
\colhead{HD} & \colhead{\Teff} & \colhead{\logg} & \colhead{\vmicro} & \colhead{[Fe/H]} & \colhead{$\sigma_{\rm [Fe/H]}$} & \colhead{$N_{\rm I}$} & \colhead{$N_{\rm II}$}
}
\startdata
166 & 5550 & 4.50 & 0.80 & 0.13 & 0.05 & 386 & 20\\
4614 & 5900 & 4.50 & 0.90 & $-$0.35 & 0.04 & 303 & 21\\
4628 & 5150 & 4.60 & 0.80 & $-$0.21 & 0.08 & 397 & 13\\
10476 & 5200 & 4.35 & 0.00 & 0.03 & 0.07 & 409 & 18\\
10700\tablenotemark{a} & 5300 & 4.40 & 0.00 & $-$0.57 & 0.05 & 361 & 13\\
12235\tablenotemark{v} & 6150 & 4.40 & 1.55 & 0.31 & 0.07 & 411 & 25\\
16160 & 5100 & 4.55 & 0.60 & $-$0.03 & 0.14 & 388 & 13\\
16895 & 6500 & 4.70 & 1.70 & 0.07 & 0.08 & 307 & 22\\
22484 & 6050 & 4.30 & 1.50 & $-$0.10 & 0.05 & 343 & 28\\
25680 & 5900 & 4.60 & 1.15 & 0.02 & 0.04 & 376 & 26\\
26965 & 5300 & 4.55 & 0.70 & $-$0.26 & 0.08 & 431 & 14\\
32147\tablenotemark{v} & 4900 & 4.30 & 0.80 & 0.38 & 0.10 & 278 & 6\\
48682 & 6150 & 4.50 & 1.35 & 0.08 & 0.06 & 416 & 24\\
50281 & 5100 & 4.50 & 1.40 & 0.00 & 0.13 & 378 & 14\\
52711 & 6000 & 4.55 & 1.25 & $-$0.13 & 0.04 & 336 & 22\\
61421\tablenotemark{b} & 6850 & 4.55 & 2.40 & $-$0.04 & 0.06 & 297 & 25\\
69830\tablenotemark{a} & 5450 & 4.40 & 0.10 & 0.04 & 0.05 & 421 & 20\\
69897 & 6450 & 4.60 & 1.75 & $-$0.31 & 0.06 & 285 & 21\\
76151\tablenotemark{v} & 5750 & 4.40 & 0.85 & 0.08 & 0.04 & 432 & 24\\
78366 & 6050 & 4.60 & 1.30 & 0.01 & 0.04 & 358 & 25\\
82328\tablenotemark{b} & 6450 & 4.20 & 2.10 & $-$0.17 & 0.07 & 255 & 24\\
84737 & 5950 & 4.30 & 1.30 & 0.07 & 0.04 & 396 & 30\\
90839 & 6300 & 4.65 & 1.40 & $-$0.10 & 0.05 & 289 & 25\\
126053 & 5650 & 4.40 & 0.65 & $-$0.44 & 0.04 & 345 & 18\\
149661\tablenotemark{v} & 5300 & 4.40 & 0.00 & 0.11 & 0.07 & 442 & 14\\
155886\tablenotemark{b} & 5100 & 4.30 & 0.00 & $-$0.30 & 0.07 & 373 & 13\\
157214 & 5650 & 4.35 & 0.60 & $-$0.47 & 0.04 & 320 & 20\\
166620 & 5200 & 4.50 & 0.50 & $-$0.09 & 0.10 & 424 & 15\\
170657 & 5200 & 4.55 & 0.60 & $-$0.11 & 0.08 & 422 & 14\\
185144 & 5400 & 4.50 & 1.00 & $-$0.20 & 0.06 & 422 & 17\\
186408 & 5780 & 4.35 & 0.85 & 0.09 & 0.04 & 375 & 25\\
201091 & 4800 & 4.35 & 0.90 & $-$0.19 & 0.15 & 238 & 3\\
215648 & 6350 & 4.35 & 1.95 & $-$0.24 & 0.07 & 283 & 18\\
219134 & 5100 & 4.40 & 0.90 & 0.10 & 0.13 & 412 & 16\\
222368 & 6300 & 4.40 & 1.70 & $-$0.11 & 0.07 & 297 & 26\\
\enddata
\tablenotetext{a}{selected from \citet{Endl:02}}
\tablenotetext{b}{selected from \citet{Walk:95}}
\tablenotetext{v}{VSL dwarf}
\end{deluxetable}

\clearpage

\begin{deluxetable}{rrrrrrrr}
\tabletypesize{\footnotesize}
\tablecaption{Best-fit parameters and iron abundances for the control dwarfs. Units are as in Table~\ref{param_tab_1}. \label{param_tab_3}}
\tablewidth{0pt}
\tablehead{
\colhead{HD} & \colhead{\Teff} & \colhead{\logg} & \colhead{\vmicro} & \colhead{[Fe/H]} & \colhead{$\sigma_{\rm [Fe/H]}$} & \colhead{$N_{\rm I}$} & \colhead{$N_{\rm II}$}
}
\startdata
5015 & 6200 & 4.45 & 1.50 & 0.07 & 0.07 & 360 & 24\\
10780 & 5350 & 4.40 & 0.00 & 0.07 & 0.06 & 417 & 16\\
25998 & 6550 & 4.95 & 2.30 & 0.34 & 0.11 & 255 & 16\\
46588 & 6300 & 4.60 & 1.50 & $-$0.10 & 0.08 & 376 & 26\\
72945 & 6350 & 4.65 & 1.50 & 0.06 & 0.05 & 353 & 27\\
72946 & 5600 & 4.40 & 0.40 & 0.15 & 0.05 & 413 & 20\\
81858 & 5950 & 4.15 & 1.45 & 0.02 & 0.05 & 409 & 27\\
82885 & 5700 & 4.55 & 1.20 & 0.38 & 0.05 & 435 & 25\\
85380 & 6100 & 4.30 & 1.50 & 0.08 & 0.06 & 404 & 28\\
88595 & 6450 & 4.55 & 1.75 & 0.09 & 0.07 & 343 & 28\\
101177 & 6000 & 4.60 & 1.20 & $-$0.20 & 0.04 & 335 & 22\\
101501 & 5600 & 4.55 & 0.75 & 0.00 & 0.04 & 420 & 22\\
101563 & 5900 & 4.10 & 1.35 & $-$0.02 & 0.06 & 382 & 24\\
103095 & 5250 & 5.00 & 0.00 & $-$1.25 & 0.11 & 257 & 4\\
109358 & 6000 & 4.55 & 1.30 & $-$0.22 & 0.05 & 358 & 20\\
110897 & 6000 & 4.60 & 1.20 & $-$0.56 & 0.05 & 282 & 16\\
111395 & 5650 & 4.50 & 0.70 & 0.11 & 0.05 & 432 & 22\\
126141 & 7100 & 4.85 & 2.25 & 0.04 & 0.08 & 275 & 21\\
143333 & 6350 & 4.40 & 1.80 & $-$0.02 & 0.09 & 308 & 21\\
176051 & 5950 & 4.50 & 1.30 & $-$0.14 & 0.05 & 311 & 24\\
179957 & 5750 & 4.40 & 0.85 & 0.00 & 0.04 & 397 & 21\\
179958 & 5750 & 4.40 & 0.75 & 0.04 & 0.04 & 377 & 24\\
184151 & 6750 & 4.35 & 2.40 & $-$0.20 & 0.10 & 221 & 21\\
187691 & 6150 & 4.40 & 1.40 & 0.10 & 0.05 & 356 & 27\\
188376 & 5550 & 3.95 & 1.00 & 0.05 & 0.05 & 384 & 25\\
193555 & 6500 & 4.60 & 2.55 & 0.50 & 0.14 & 198 & 14\\
193664 & 6000 & 4.50 & 1.15 & $-$0.10 & 0.04 & 333 & 24\\
197963 & 6450 & 4.40 & 1.85 & 0.13 & 0.07 & 316 & 25\\
216172 & 7000 & 4.95 & 2.85 & 0.49 & 0.14 & 184 & 16\\
216172 & 7000 & 5.00 & 2.40 & 0.49 & 0.12 & 229 & 17\\
221445 & 6400 & 4.40 & 1.90 & $-$0.14 & 0.03 & 194 & 16\\
224930 & 5400 & 4.30 & 0.00 & $-$0.90 & 0.07 & 304 & 13\\
\enddata
\end{deluxetable}

\clearpage

\begin{deluxetable}{rrrrrrrr}
\tabletypesize{\footnotesize}
\tablecaption{Best-fit parameters and iron abundances for the VSL dwarfs not listed in Tables~\ref{param_tab_1} to \ref{param_tab_3}. Units are as in Table~\ref{param_tab_1}. \label{param_tab_4}}
\tablewidth{0pt}
\tablehead{
\colhead{HD} & \colhead{\Teff} & \colhead{\logg} & \colhead{\vmicro} & \colhead{[Fe/H]} & \colhead{$\sigma_{\rm [Fe/H]}$} & \colhead{$N_{\rm I}$} & \colhead{$N_{\rm II}$}
}
\startdata
1461 & 5700 & 4.30 & 0.50 & 0.20 & 0.05 & 399 & 22\\
9562 & 5850 & 4.20 & 1.20 & 0.18 & 0.06 & 423 & 25\\
20630 & 5750 & 4.55 & 1.15 & 0.05 & 0.05 & 386 & 21\\
73752 & 5900 & 4.35 & 1.35 & 0.39 & 0.08 & 407 & 21\\
99491 & 5650 & 4.45 & 1.00 & 0.40 & 0.06 & 443 & 23\\
99492 & 5250 & 4.60 & 1.20 & 0.36 & 0.12 & 415 & 16\\
104304 & 5750 & 4.45 & 1.15 & 0.35 & 0.06 & 362 & 20\\
114710 & 6050 & 4.55 & 1.20 & 0.02 & 0.05 & 313 & 23\\
115617 & 5600 & 4.35 & 0.60 & 0.03 & 0.04 & 345 & 21\\
125184 & 5800 & 4.25 & 1.20 & 0.38 & 0.05 & 352 & 23\\
135101 & 5700 & 4.20 & 0.90 & 0.06 & 0.04 & 430 & 24\\
135101 & 5550 & 4.25 & 0.40 & 0.10 & 0.05 & 424 & 24\\
182572 & 5750 & 4.35 & 1.40 & 0.45 & 0.08 & 351 & 24\\
\enddata
\end{deluxetable}

\clearpage

\begin{deluxetable}{cccp{5mm}ccc}
\tablecaption{Best-fit parameters for HD\,68988 using different parameters for the determination of the solar reference abundances. \label{reference}}
\tablewidth{0pt}
\tablehead{
\multicolumn{3}{c}{Reference model} & & \multicolumn{3}{c}{HD\,68988} \\ 
\colhead{\Teff} & \colhead{\logg}  & \colhead{\vmicro} & & \colhead{\Teff} & \colhead{\logg}  & \colhead{\vmicro}
}
\startdata
5720 & 4.35 & 0.70 & & 5950 & 4.35 & 1.25 \\
5720 & 4.35 & 0.90 & & 5950 & 4.30 & 1.40 \\
5720 & 4.55 & 0.70 & & 5950 & 4.55 & 1.35 \\
5720 & 4.55 & 0.90 & & 5950 & 4.55 & 1.45 \\
5840 & 4.35 & 0.70 & & 6050 & 4.35 & 1.25 \\
5840 & 4.35 & 0.90 & & 6050 & 4.35 & 1.40 \\
5840 & 4.55 & 0.70 & & 6050 & 4.55 & 1.30 \\
5840 & 4.55 & 0.90 & & 6050 & 4.55 & 1.45 \\
\enddata
\end{deluxetable}

\clearpage

\begin{deluxetable}{rcccccc}
\tablecaption{Iron abundances derived when using parameters from \citet{Felt:01,Fuhr:98,Take:02b,Edva:93a,Gonz:01,Sant:03} instead of that given in Tables~\ref{param_tab_1}, \ref{param_tab_3} and \ref{param_tab_4}, and for HD\,125184 and HD\,177830 also using the subset of lines given in the cited works (additional rows). $\Delta_{\rm ref}$ = [Fe](this work) $-$ [Fe](reference), $\Delta_{\rm ion}$ = [Fe I] $-$ [Fe II]. \label{diff}}
\tablewidth{0pt}
\tablehead{
\colhead{HD} & \colhead{\Teff} & \colhead{\logg} & \colhead{\vmicro} & \colhead{N(I/II)} & \colhead{$\Delta_{\rm ref}$} & \colhead{$\Delta_{\rm ion}$}
}
\startdata
  99491    & 5300 & 4.12 & 1.00  & 425/23   & +0.10 & $-$0.15 \\
 103095    & 5020 & 4.61 & 0.67  & 256/4    & +0.04 & $-$0.06 \\
 104304    & 5400 & 4.12 & 1.15  & 342/22   & +0.05 & $-$0.21 \\
 125184    & 5562 & 3.92 & 1.65  & 333/23   & +0.05 & $-$0.08 \\
               &      &      &       &  22/1    & $-$0.06 & $-$0.06 \\
 177830    & 4829 & 3.46 & 1.08  & 270/7    & +0.22 & $-$0.17 \\
           & 4818 & 3.32 & 0.97  &  21/3    & +0.10 & $-$0.02 \\
\enddata
\end{deluxetable}

\clearpage

\begin{deluxetable}{lcrrcrcrcr}
\tabletypesize{\footnotesize}
\tablecaption{Mean and standard deviations of the metallicities and temperatures for all stars in each sample (columns 2--4). The remaining columns list mean metallicities for stars contained in bins of 200~K centered on the indicated temperature.\label{stat_tab}}
\tablewidth{0pt}
\tablehead{
\colhead{Name} & \colhead{N} & \colhead{[Fe/H]} & \colhead{\Teff [K]} & \colhead{N}  & \colhead{[Fe/H]$_{5700}$} & \colhead{N} & \colhead{[Fe/H]$_{5900}$} & \colhead{N} & \colhead{[Fe/H]$_{6100}$} 
}
\startdata
CGP dwarfs     & 34 &    +0.14 $\pm$ 0.26 & 5872 $\pm$ 336 & 8 & +0.30 $\pm$ 0.17 & 6 & +0.05 $\pm$ 0.20 & 9 & +0.00 $\pm$ 0.34 \\
CGP dwarfs\tablenotemark{1} & 28 & +0.19 $\pm$ 0.21 & 5846 $\pm$ 354 & 7 & +0.34 $\pm$ 0.11 & 5 & +0.03 $\pm$ 0.22 & 6 & +0.11 $\pm$ 0.17 \\
CGP dwarfs\tablenotemark{2} & 22 & +0.21 $\pm$ 0.17 & 5889 $\pm$ 323 & 6 & +0.35 $\pm$ 0.12 & 3 & +0.12 $\pm$ 0.21 & 6 & +0.11 $\pm$ 0.17 \\
no-CGP dwarfs  & 35 &  $-$0.08 $\pm$ 0.21 & 5696 $\pm$ 535 & 3 & $-$0.28 $\pm$ 0.31 & 4 & $-$0.04 $\pm$ 0.21 & 5 & +0.03 $\pm$ 0.18 \\
control dwarfs & 32 &  $-$0.02 $\pm$ 0.36 & 6095 $\pm$ 483 & 6 & +0.11 $\pm$ 0.14 & 3 & $-$0.05 $\pm$ 0.08 & 6 & $-$0.15 $\pm$ 0.24 \\
CGP + no-CGP   & 69 &    +0.03 $\pm$ 0.26 & 5783 $\pm$ 454 & & & & & & \\
VSL dwarfs     & 21 &    +0.26 $\pm$ 0.17 & 5664 $\pm$ 271 & & & & & & \\
\enddata
\tablenotetext{1}{Excluding stars with companions with $M\sin i \ge 6.5~M_{\rm J}$}
\tablenotetext{2}{Additionally excluding stars where companion might be brown dwarf candidate or M dwarf according to \citet{Han:01}}
\end{deluxetable}

\clearpage

\begin{deluxetable}{rccrrrp{3mm}ccrrr}
\tabletypesize{\footnotesize}
\tablecaption{Observing log for spectra of additional stars. For column explanation see Table~\ref{obs_log_dwp}. \label{obs_log_a}}
\tablewidth{0pt}
\tablehead{
 & \multicolumn{5}{c}{blue spectrum} & & \multicolumn{5}{c}{red spectrum} \\ 
\colhead{HD} & \colhead{$\lambda_{\rm c}$ [\AA]}  & \colhead{Date} & \colhead{$t_{\rm i}$ [s]} & \colhead{S/N}  & \colhead{O} &  &  \colhead{$\lambda_{\rm c}$ [\AA]}  & \colhead{Date} & \colhead{$t_{\rm i}$ [s]} & \colhead{S/N}  & \colhead{O}
}
\startdata
16141      &  5136 &   2000-08-21 &  1800 &   205 &    L &  &  6115 &   2000-08-18 &   900 &   236 &    L\\
18445      &  5239 &   1999-10-20 &  2700 &   212 &    L &  &  6211 &   1999-10-23 &  1800 &   332 &    L\\
 &  5239 &   1999-10-20 &  1800 &   204 &    L &  &  6211 &   1999-10-23 &  1200 &   266 &    L\\
22049      &  5240 &   1999-10-19 &    75 &   247 &    L &  &  6212 &   1999-10-24 &    45 &   347 &    L\\
29587      &  5239 &   1999-10-19 &  1800 &   272 &    L &  &  6210 &   1999-10-23 &   900 &   292 &    L\\
37124      &  5240 &   2000-01-29 &  1800 &   202 &    L &  &  6263 &   2000-01-26 &  1200 &   227 &    L\\
74156      &  5276 &   2001-10-25 &  1800 &   188 &    L &  &  6263 &   2001-10-28 &  2700 &   170 &    L\\
80606      &  5276 &   2002-01-23 &  2700 &   149 &    H &  &  6263 &   2002-01-28 &  2700 &   177 &    H\\
80607      &  5276 &   2002-01-23 &  2700 &   108 &    H &  &  6263 &   2002-01-28 &  2700 &   174 &    H\\
82943      &  5276 &   2002-01-22 &  1200 &   207 &    H &  &  6263 &   2002-01-25 &  1800 &   275 &    H\\
89744      &  5204 &   2001-05-08 &   900 &   231 &    L &  &  6065 &   2001-05-14 &   750 &   381 &    L\\
95128      &  5135 &   1999-05-30 &   240 &   269 &    L &  &  6017 &   1999-05-26 &   300 &   337 &    L\\
 & & & & &  &  &  6263 &   2000-01-25 &   200 &   282 &    L\\
106252     &  5204 &   2001-05-08 &  2700 &   236 &    L &  &  6065 &   2001-05-14 &  1800 &   217 &    L\\
108874     &  5241 &   2003-02-18 &  2700 &   117 &    H  & &  6214 &   2003-02-22 &  2700 &   175 &    H\\
110833     &  5240 &   2000-01-29 &  1800 &   166 &    L &  &  6263 &   2000-01-25 &   900 &   286 &    L\\
112758     &  5135 &   1999-05-30 &  1800 &   177 &    L &  &  6211 &   1999-05-27 &  1800 &   271 &    L\\
140913     &  5135 &   1999-05-28 &  2700 &   173 &    L &  &  6212 &   1999-05-27 &  2700 &   233 &    L\\
168443     &  5136 &   1999-05-30 &  1800 &   229 &    L &  &  6213 &   1999-10-25 &   900 &   323 &    L\\
168746     &  5135 &   2000-08-21 &  1800 &   173 &    L &  &  6113 &   2000-08-18 &  1800 &   191 &    L\\
186427     &  5136 &   1999-05-28 &   900 &   196 &    L &  &  6019 &   1999-05-26 &   900 &   379 &    L\\
 & & & & &  &  &  6114 &   2000-08-19 &   360 &   264 &    L\\
192263     &  5135 &   2000-08-21 &  2100 &   184 &    L &  &  6113 &   2000-08-19 &  1800 &   249 &    L\\
197964     &  5135 &   2000-08-22 &   180 &   243 &    L &  &  6113 &   2000-08-18 &    90 &   272 &    L\\
202206     &  5204 &   2001-08-23 &  2700 &   154 &    L &  &  6212 &   2001-08-28 &  1800 &   191 &    L\\
210277     &  5240 &   1999-10-19 &   750 &   199 &    L &  &  6212 &   1999-10-25 &   900 &   358 &    L\\
222582     &  5135 &   2000-08-21 &  1800 &   153 &    L &  &  6113 &   2000-08-18 &  1009 &   164 &    L\\
BD--040782  &  5240 &   1999-10-20 &  2700 &   116 &    L &  &  6162 &   1999-10-23 &  2700 &   229 &    L\\
BD--103166  &  5239 &   2001-05-10 &  2700 &    89 &    L & &  6263 &   2002-01-26 &  2700 &    52 &     H\\
& & & & & &  &  6263 &   2002-01-28 &  2700 &    75 & H  \\
\enddata
\end{deluxetable}

\clearpage

\begin{deluxetable}{crrrrrrrr}
\tabletypesize{\footnotesize}
\tablecaption{Best-fit parameters and iron abundances for additional stars. They are dwarfs with planets that have either low mass or high eccentricity or both, except for those noted in the column marked with $\ast$: a\dots companion of HD\,80606 (planet status unknown), b\dots stars with a brown dwarf companion \citep[$M\sin i \ge 13~M_{\rm J}$,][]{Zuck:01}, c\dots planet status controversial, g\dots giant star originally included in the control dwarf list, m\dots stars that have been added to the planet search programs based on their high metallicities. Units are as in Table~\ref{param_tab_1}. \label{param_tab_a}}
\tablewidth{0pt}
\tablehead{
\colhead{$\ast$} & \colhead{HD} & \colhead{\Teff} & \colhead{\logg} & \colhead{\vmicro} & \colhead{[Fe/H]} & \colhead{$\sigma_{\rm [Fe/H]}$} & \colhead{$N_{\rm I}$} & \colhead{$N_{\rm II}$}
}
\startdata
  & 16141 & 5750 & 4.20 & 0.90 & 0.16 & 0.05 & 383 & 28\\
b & 18445 & 5300 & 4.60 & 1.40 & 0.01 & 0.11 & 368 & 14\\
  & 22049 & 5200 & 4.50 & 0.60 & $-$0.01 & 0.08 & 467 & 18\\
b & 29587 & 5700 & 4.55 & 0.80 & $-$0.62 & 0.04 & 300 & 15\\
  & 37124 & 5700 & 4.50 & 1.10 & $-$0.38 & 0.05 & 379 & 17\\
  & 74156 & 6050 & 4.35 & 1.25 & 0.09 & 0.05 & 392 & 25\\
  & 80606 & 5700 & 4.40 & 0.90 & 0.46 & 0.07 & 445 & 23\\
a & 80607 & 5700 & 4.40 & 1.00 & 0.45 & 0.08 & 457 & 27\\
  & 82943\tablenotemark{h} & 5900 & 4.40 & 0.75 & 0.24 & 0.04 & 403 & 26\\
  & 89744\tablenotemark{h} & 6300 & 4.40 & 1.80 & 0.22 & 0.08 & 337 & 27\\
  & 95128 & 6000 & 4.40 & 1.30 & 0.00 & 0.04 & 381 & 25\\
  & 106252 & 5900 & 4.40 & 1.10 & $-$0.10 & 0.04 & 349 & 25\\
m & 108874 & 5750 & 4.45 & 1.15 & 0.31 & 0.05 & 405 & 24\\
b & 110833 & 5200 & 4.40 & 0.80 & 0.16 & 0.10 & 445 & 19\\
b & 112758 & 5300 & 4.60 & 0.80 & $-$0.39 & 0.07 & 353 & 13\\
b & 140913 & 6100 & 4.60 & 1.50 & 0.09 & 0.07 & 340 & 20\\
  & 168443 & 5600 & 4.10 & 0.80 & 0.09 & 0.04 & 378 & 20\\
  & 168746 & 5600 & 4.35 & 0.60 & $-$0.04 & 0.04 & 366 & 21\\
  & 186427 & 5800 & 4.40 & 0.95 & 0.06 & 0.03 & 356 & 26\\
c & 192263 & 5100 & 4.45 & 0.00 & 0.11 & 0.10 & 386 & 16\\
g & 197964 & 5050 & 3.35 & 1.55 & 0.31 & 0.11 & 395 & 24\\
b & 202206 & 5800 & 4.50 & 1.00 & 0.40 & 0.05 & 417 & 20\\
  & 210277 & 5700 & 4.40 & 1.20 & 0.27 & 0.06 & 405 & 23\\
  & 222582\tablenotemark{h} & 5800 & 4.40 & 0.80 & 0.03 & 0.04 & 366 & 24\\
b & BD--04 0782 & 5200 & 4.50 & 2.00 & 0.05 & 0.20 & 290 & 8\\
m & BD--10 3166 & 5550 & 4.45 & 1.00 & 0.51 & 0.10 & 447 & 23\\
\enddata
\tablenotetext{h}{Companion might be brown dwarf candidate according to \citet[ their group 2]{Han:01}}
\end{deluxetable}

\end{document}